\documentclass[twocolumn]{aastex631}

\usepackage{tikz}
\usetikzlibrary{shapes,arrows}
\usepackage{graphicx}
\usepackage{amssymb}
\usepackage{subfigure}
\usepackage{epstopdf}
\usepackage{gensymb}
\usepackage{mathtools}
\usepackage{commath}
\usepackage{diagbox}
\usepackage{makecell}
\usepackage{CJKutf8} 
\usepackage{tabularx}

\newcommand{\pccc}{\ensuremath{{\rm pc}\,{\rm cm}^{-3}}}

\graphicspath{{./}{figures/}}

\begin{document}

\title{Calculation and Uncertainty of Fast Radio Burst Structure Based on Smoothed Data 
\footnote{Draft, \today}}

\author[0000-0002-9116-307X]{Adrian~T.~Sutinjo}
\affiliation{International Centre for Radio Astronomy Research, Curtin University, Bentley, WA 6102, Australia}

\author[0000-0002-6895-4156]{Danica~R.~Scott}
\affiliation{International Centre for Radio Astronomy Research, Curtin University, Bentley, WA 6102, Australia}

\author[0000-0002-6437-6176]{Clancy~W.~James}
\affiliation{International Centre for Radio Astronomy Research, Curtin University, Bentley, WA 6102, Australia}

\author[0000-0002-5067-8894]{Marcin Glowacki}
\affiliation{International Centre for Radio Astronomy Research, Curtin University, Bentley, WA 6102, Australia}

\author[0000-0003-2149-0363]{Keith W. Bannister}
\affiliation{ATNF, CSIRO, Space and Astronomy, PO Box 76, Epping, NSW 1710, Australia}

\author[0000-0002-2858-9481]{Hyerin Cho \begin{CJK}{UTF8}{}
 \CJKfamily{mj}(조혜린)\end{CJK}}
\affiliation{Center for Astrophysics $\vert$ Harvard \& Smithsonian, 60 Garden Street, Cambridge, MA 02138, USA}
\affiliation{Black Hole Initiative at Harvard University, 20 Garden Street, Cambridge, MA 02138, USA}

\author[0000-0002-8101-3027]{Cherie~K.~Day}
\affiliation{Centre for Astrophysics and Supercomputing, Swinburne University of Technology, Hawthorn, VIC, 3122, Australia}

\author[0000-0001-9434-3837]{Adam~T.~Deller}
\affiliation{Centre for Astrophysics and Supercomputing, Swinburne University of Technology, Hawthorn, VIC, 3122, Australia}

\author{Timothy~P.~Perrett}
\affiliation{International Centre for Radio Astronomy Research, Curtin University, Bentley, WA 6102, Australia}

\author[0000-0002-7285-6348]{Ryan~M.~Shannon}
\affiliation{Centre for Astrophysics and Supercomputing, Swinburne University of Technology, Hawthorn, VIC, 3122, Australia}

\begin{abstract}
Studies of the time-domain structure of fast radio bursts (FRBs) require an accurate estimate of the FRB dispersion measure in order to recover the intrinsic burst shape. Furthermore, the exact DM is itself of interest when studying the time-evolution of the medium through which multiple bursts from repeating FRBs propagate.
A commonly used approach to obtain the
dispersion measure is to take the value that maximizes the FRB structure in the time domain.
However, various authors use differing methods to obtain this structure parameter, and do not document the smoothing method used. Furthermore, there are no quantitative estimates of the error in this procedure in the FRB literature. In this letter, we present a smoothing filter based on the discrete cosine transform, and show that computing the structure parameter by summing the squares of the derivatives and taking the square root (that is, the 2-norm,  $\sqrt{\Sigma (d/dt)^2}$) immediately lends itself to calculation of uncertainty of the structure parameter. We illustrate this with FRB181112 and FRB210117 data, which were detected by the Australian Square Kilometre Array Pathfinder, and for which high-time-resolution data is available.

\end{abstract}

\keywords{
Radio transient sources (2008)---Astronomy data analysis (1858)
}

\section{Introduction} \label{sec:intro}

Fast radio bursts (FRBs) are extragalactic radio transients of millisecond duration~\citep{Lorimer2007}. Both the underlying mechanism of their emission, and the properties of the astrophysical plasmas through which they propagate, can be studied through detailed analysis of their time-frequency structure \citep[e.g.][]{Michilli2018_121102,Macquart2020}. The frequency dependent delay (dispersion measure, DM) of FRBs informs about the total integrated column density of free electrons --- and hence total matter density --- through which the FRB has propagated. Long-term studies of DM variation in repeating FRBs yield information on the medium in the vicinity of the progenitor \citep[e.g.][]{Zhao2021_DM_evolution}. Additionally, studies of FRB structure are sensitive to the assumed value of DM used to dedisperse the FRB \citep[e.g.][]{Hessels_2019}.
To this end, FRB analysis aims to identify the correct DM, i.e.\ that corresponding to the integrated electron column density through which the FRB has passed. While FRB searches use
the dispersion measure that maximizes signal-to-noise ratio (S/N),
it is often assumed that the correct DM is that which maximizes structure~\citep{Caleb_10.1093/mnras/staa1791, Hilmarsson_10.1093/mnras/stab2936, 2021MNRAS.505.3041P}.

A common measure of structure is the time-derivative of the pulse intensity $I(t)$~\citep{Gajjar_2018, Hessels_2019, Josephy_2019, Andersen2019, Pilia_2020}. However, since $I(t)$ is noisy, the derivative operation is not applied to a noisy estimate of the $I(t)$, but rather to a smoothed version thereof.
This smoothing operation has the potential to affect the structure-maximizing DM, since over-smoothing will smear out the intrinsic FRB structure, while under-smoothing will leave the derivative dominated by noise.  However, to date, there has been little attention paid on how to optimally select a smoothing time.

In the literature, there is also a slight variation in how the resulting $d/dt$ are combined. For example, \citet{Hessels_2019} used $\Sigma \left(d/dt\right)^2$, \citet{Gajjar_2018} suggested $\Sigma \left|d/dt\right|$, and \citet{Josephy_2019} experimented with $\Sigma \left(d/dt\right)^4$. One may argue that this is a matter of preference as long as the positive of the $d/dt$ is combined and an interesting structure is detected.
However, while any of these approaches should be able to identify the DM that maximises the structure in the signal, the uncertainty of the resultant fitted DM is also of considerable interest --- without an accurate estimate of this uncertainty, it is impossible to determine whether, for instance, the DM of a given source is changing between bursts.

In \secref{sec:method}, we shall demonstrate that the vector norm offers a meaningful measure of the uncertainty in the structure parameter inferred from noisy time-series data. In particular, the  $\sqrt{\Sigma (d/dt)^2}$ (Euclidean or 2-norm) offers a physical and intuitive interpretation, as illustrated in \secref{sec:result}. \secref{sec:apply} then illustrates how the structure parameter, and uncertainty thereon, to be calculated for two example FRBs. Under the assumption that the ``correct'' DM of an FRB\footnote{i.e.\ such that ${\rm DM}=\int n(e) (1+z)^{-1} d\ell$ for electron density $n(e)$ and unique signal propagation path $d \ell$.} is synonymous with structure maximization, this allows the correct DM to be identified, and assigned an uncertainty 
$\pm \Delta \text{DM}$ which quantifies the confidence in such a detection.

\section{Method} \label{sec:method}
\subsection{Matrix representation and uncertainty definition} \label{sec:unc_def}
Let $\mathbf{i}=[I(1),\cdots,I(N)]^T$ be a vector that contains discrete time samples of noisy total intensity $I(t)$. We may represent the smoothing operator as a matrix $\mathbf{S}$ and the first derivative as a matrix $\mathbf{D}_1$. Our immediate discussion here pertains to any $\mathbf{S}$ and $\mathbf{D}_1$ matrices, therefore the conclusions are generally applicable.  It suffices to say, as we shall demonstrate later, it is indeed possible to write the smoothing algorithm as a matrix; for example, see~\citet{Eilers_doi:10.1021/ac034173t, STICKEL2010467, Strang_wavelet}. The result of the smoothing process is the smoothed estimate
\begin{eqnarray}
\tilde{\mathbf{i}}=\mathbf{S}\mathbf{i}.
\label{eqn:smoothed}
\end{eqnarray}
The first derivative operator can be expressed as a matrix, for example, a simple forward difference on $N$ data points is a $(N-1) \times N$ matrix
\begin{eqnarray}
\mathbf{D}_1=
\left[ \begin{array}{c c c c} 
-1 & 1 &  &  \\ 
 & \ddots & \ddots & \\
   &   & -1 & ~1
\end{array} \right].
\label{eqn:D1}
\end{eqnarray}
Therefore, the first derivative of the smoothed data is 
\begin{eqnarray}
\mathbf{D}_1\tilde{\mathbf{i}}=\mathbf{D}_1\mathbf{S}\mathbf{i}.
\label{eqn:diff_smoothed}
\end{eqnarray}

\begin{figure}
    \centering
    \begin{tikzpicture}
\draw [thick, ->] (0,0) -- (2,1);
\node at (0.8,0.8) {$\mathbf{D}_1\mathbf{S}\mathbf{\tilde{i}}$};
\draw [gray, thick, ->] (2,1) -- (2.707,1);
\node at (3.4,1) {$\mathbf{D}_1\mathbf{S}\mathbf{\Delta_i}$};
\draw [gray, dashed] (2,1) circle [radius=0.707];
\draw [thick, ->] (0,0) -- (2.7,0.95);
\node at (1.2,0.2) {$\mathbf{D}_1\mathbf{\tilde{i}}$};
   \end{tikzpicture}
    \caption{Vector diagram representing Eq.~\eqref{eqn:D1_smoothed}.}
    \label{fig:vec_sum}
\end{figure}
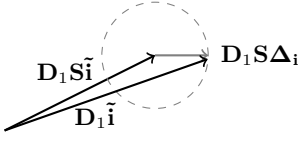

The original noisy data can be expressed as the sum of the smoothed estimate and the difference between the noisy data and the smoothed version $\mathbf{\Delta_i}=\mathbf{i}-\tilde{\mathbf{i}}$. 
\begin{eqnarray}
\mathbf{i}=\tilde{\mathbf{i}}+\mathbf{\Delta_i}.
\label{eqn:i_Del}
\end{eqnarray}
Therefore, Eq.~\eqref{eqn:diff_smoothed} may be written as
\begin{eqnarray}
\mathbf{D}_1\tilde{\mathbf{i}}&=&\mathbf{D}_1\mathbf{S}(\tilde{\mathbf{i}}+\mathbf{\Delta_i})\nonumber\\
&=&\mathbf{D}_1\mathbf{S}\tilde{\mathbf{i}}+\mathbf{D}_1\mathbf{S}\mathbf{\Delta_i}.
\label{eqn:D1_smoothed}
\end{eqnarray}
Eq.~\eqref{eqn:D1_smoothed} is illustrated in Fig.~\ref{fig:vec_sum}. We note that the first term in the right hand side of Eq.~\eqref{eqn:D1_smoothed}, $\mathbf{D}_1\mathbf{S}\tilde{\mathbf{i}}=\mathbf{D}_1\mathbf{S}\mathbf{S}\mathbf{i}$, is the derivative of the ``double-smoothed'' data. The second term, $\mathbf{D}_1\mathbf{S}\mathbf{\Delta_i}$, is the derivative of the smoothed ``de-trended'' noise. 

The length (2-norm, $\norm{.}_2$) of $\mathbf{D}_1\mathbf{S}\mathbf{\Delta_i}$ is the radius of the dashed circle in Fig.~\ref{fig:vec_sum}, which may be written as 
\begin{eqnarray}
\norm{\mathbf{D}_1\mathbf{S}\mathbf{\Delta_i}}_2=\sqrt{\sum_{n=1}^N \left[(\mathbf{D}_1\mathbf{S}\mathbf{i})_n-(\mathbf{D}_1\mathbf{S}\tilde{\mathbf{i}})_n\right]^2},
\label{eqn:norm_D1Si}
\end{eqnarray}
where $(.)_n$ indicates the $n$-th element of the vector in question. The right hand side of Eq.~\eqref{eqn:norm_D1Si} suggests that it can be interpreted as proportional to a standard deviation estimate ($\tilde{\sigma}\sqrt{N-1}$) of the first derivative of the smoothed data. Taking this as the uncertainty estimate of $\mathbf{D}_1\tilde{\mathbf{i}}$ means we take $\mathbf{D}_1\mathbf{S}\mathbf{\Delta_i}$ to represent a particular noise instance with standard deviation of $\norm{\mathbf{D}_1\mathbf{S}\mathbf{\Delta_i}}_2/\sqrt{N-1}$. In that case, the tip of the vector $\mathbf{D}_1\tilde{\mathbf{i}}$  in Fig.~\ref{fig:vec_sum} could lie anywhere on the circumference of the grey circle such that the estimate of relative uncertainty may be computed as
\begin{eqnarray}
\frac{\norm{\mathbf{D}_1\mathbf{S}\mathbf{\Delta_i}}_2}{\norm{\mathbf{D}_1\mathbf{S}\tilde{\mathbf{i}}}_2}.
\label{eqn:rel_error}
\end{eqnarray}
The structure parameter that is consistent with this uncertainty estimate is, evidently, the 2-norm of the $\mathbf{D}_1\tilde{\mathbf{i}}$ since  
\begin{eqnarray}
\norm{\mathbf{D}_1\mathbf{S}\tilde{\mathbf{i}}}-\norm{\mathbf{D}_1\mathbf{S}\mathbf{\Delta_i}}\leq\norm{\mathbf{D}_1\tilde{\mathbf{i}}}\leq\norm{\mathbf{D}_1\mathbf{S}\tilde{\mathbf{i}}}+\norm{\mathbf{D}_1\mathbf{S}\mathbf{\Delta_i}},
\label{eqn:D1_norm}
\end{eqnarray}
where the subscript $._2$ is not shown for brevity. Eq.~\eqref{eqn:D1_norm} is valid for the 2-norm for N-dimensional vectors. It can be shown formally by applying the inner product definition $\mathbf{x}^T\mathbf{y}=\norm{\mathbf{x}}_2\norm{\mathbf{y}}_2  \cos\theta$ in the expression $\norm{\mathbf{x} + \mathbf{y}}_2^2=(\mathbf{x} + \mathbf{y})^T(\mathbf{x} + \mathbf{y})$~\citep[see Ch.~1]{Strang_LALD2019} and realizing that $-1\leq \cos\theta \leq 1$ for real vectors. 

In reality, the data is a vector with N entries. In that case, the circumference of the circle in Fig.~\ref{fig:vec_sum} becomes the surface of an N-dimensional sphere with the same radius. The conclusion is unchanged.

Finally, the reciprocal of Eq.~\eqref{eqn:rel_error} is the structure to the structure noise ratio.
\begin{eqnarray}
\text{SSNR}=\frac{\norm{\mathbf{D}_1\mathbf{S}\tilde{\mathbf{i}}}_2}{\norm{\mathbf{D}_1\mathbf{S}\mathbf{\Delta_i}}_2}.
\label{eqn:SSNR}
\end{eqnarray}
This quantity may offer a more prominent visual contrast in some cases. 

For some applications, it may be desirable to test the hypothesis of whether the intrinsic structure in a signal ${\mathbf{i}_1}$ is significantly greater than that in another signal ${\mathbf{i}_2}$. This test asks whether or not the measured difference in structure,
\begin{eqnarray}
\norm{\mathbf{D}_1\mathbf{S} {\mathbf{i}_1}} - \norm{\mathbf{D}_1\mathbf{S} {\mathbf{i}_2}}, \label{eqn:rel_structure}
\end{eqnarray}
is significant compared to the relative noise,
\begin{eqnarray}
\norm{\mathbf{D}_1\mathbf{S} \mathbf{\Delta}_{\mathbf{i}_1} - \mathbf{D}_1\mathbf{S} \mathbf{\Delta}_{\mathbf{i}_2} } = \norm{\mathbf{D}_1\mathbf{S}\left( \mathbf{\Delta}_{\mathbf{i}_1} - \mathbf{\Delta}_{\mathbf{i}_2} \right) }. \label{eqn:rel_dmerror}
\end{eqnarray}
Continuing the analogy of Fig.~\ref{fig:vec_sum}, Eq.~\eqref{eqn:rel_dmerror} represents the surface of an n-dimensional sphere with magnitude given by a combination of $\mathbf{D}_1\mathbf{S} \mathbf{\Delta}_{\mathbf{i}_1}$ and $\mathbf{D}_1\mathbf{S} \mathbf{\Delta}_{\mathbf{i}_2}$. If the difference in structure between $\mathbf{i}_1$ and $\mathbf{i}_2$ given by Eq.~\eqref{eqn:rel_structure} is less than the radius of the sphere, the structures are consistent to within noise fluctuations. Specifically, the structure in $\mathbf{i}_1$ can be said to be greater than that in $\mathbf{i}_2$ when 
\begin{eqnarray}
\norm{\mathbf{D}_1\mathbf{S} {\mathbf{i}_1}} & > & \norm{\mathbf{D}_1\mathbf{S} {\mathbf{i}_2}} + \norm{\mathbf{D}_1\mathbf{S}\left( \mathbf{\Delta}_{\mathbf{i}_1} - \mathbf{\Delta}_{\mathbf{i}_2} \right) }. \label{eqn:add_error}
\end{eqnarray}

Such a test will become relevant when assessing whether the structure at one DM is significantly different from the structure at another, since $\mathbf{\Delta}_{\mathbf{i}_1}$ and $\mathbf{\Delta}_{\mathbf{i}_2}$ will in this case be correlated.

\subsection{Example smoothing matrix based on discrete cosine}\label{sec:smooth_DCT}
We can choose the eigenvectors of $\mathbf{D}_1^T\mathbf{D}_1$ as the basis for the smoothing operation. As we will show later, the close connection to the first derivative $\mathbf{D}_1$ matrix lends itself to insight when expressing the entire calculation as a single matrix operation. The eigenvectors of
\begin{eqnarray}
\mathbf{D}_1^T\mathbf{D}_1=
\left[ \begin{array}{c c c c c} 
-1 & ~1 &  &  & \\ 
-1 & ~2 & -1 &  &\\
 & . & . & . & \\
 & & -1 & 2 & -1 \\
 &   &   & -1 & ~~1 
\end{array} \right],
\label{eqn:D1TD1}
\end{eqnarray} 
are discrete cosines~\citep{Strang_DCT}. We shall call this eigenvector matrix $\mathbf{C}$ whose components are
\begin{eqnarray}
\mathbf{c}_k(j)=\cos\left(j-\frac{1}{2}\right)\frac{(k-1)\pi}{N},
\label{eqn:evec_DCT2}
\end{eqnarray} 
where $k=1,\cdots,N$ is the column number of $\mathbf{C}$ and $j=1,\cdots,N$ is the row number. The eigenvalues are
\begin{eqnarray}
\text{eig}_k=2-2\cos\frac{(k-1)\pi}{N}.
\label{eqn:eval_DCT2}
\end{eqnarray}
This is known as the discrete cosine transform-2 (DCT-2). Most importantly, DCT-2 has a fast implementation based on the fast Fourier transform (FFT)~\citep{Strang_DCT}. Therefore, numerical linear algebra eigendecomposition is not needed to obtain Eq.~\eqref{eqn:evec_DCT2} and Eq.~\eqref{eqn:eval_DCT2}. 

The smoothing operation may be written as 
\begin{eqnarray}
\tilde{\mathbf{i}}=\mathbf{C}\mathbf{D}_{\rm filter}\mathbf{C}^T\mathbf{i}=\mathbf{S}\mathbf{i}, 
\label{eqn:smooth_DCT}
\end{eqnarray}
where $\mathbf{D}_{\rm filter}$ is a diagonal matrix representing a low-pass filter in the spectral domain; $\mathbf{C}^T\mathbf{i}$ is the DCT of $\mathbf{i}$ and  $\mathbf{C}(\mathbf{D}_{\rm filter}\mathbf{C}^T\mathbf{i})$ is the inverse DCT of the filtered spectrum. The diagonal of $\mathbf{D}_{\rm filter}$ is the spectral response of a low-pass filter, for example, 
\begin{eqnarray}
f(k)=\frac{1}{1+\left(\frac{k}{k_c}\right)^{2O}},
\label{eqn:butterworth}
\end{eqnarray}
where $k_c$ is the spectral cutoff and $O$ is the order of the filter. The resulting pass band is flat and the roll off rate is $2O$ orders of magnitude per decade, as inspired by the analog Butterworth filter. The DCT is purely real and does not force a periodic boundary condition (as in the case of FFT) that could introduce a discontinuity corresponding to high-frequency contents~\citep[see Ch.~8]{Strang_wavelet}.

\begin{figure}[htb]
\begin{center}
\noindent
  \includegraphics[width=2.5in]{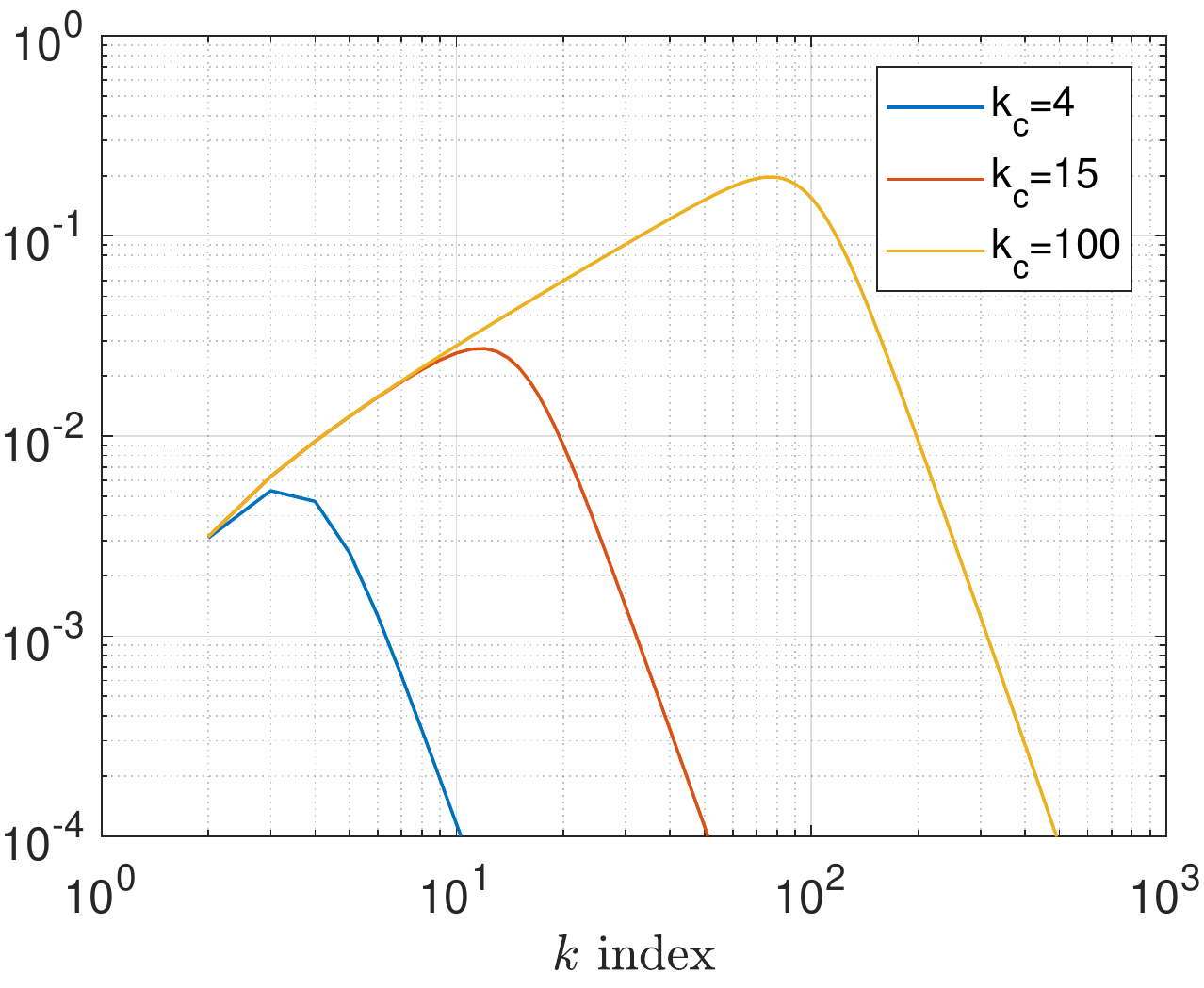}
\caption{$\mathbf{\Lambda}_{D_1^TD_1}^{1/2}\mathbf{D}_{\rm filter}$ band pass filter response for $k_c=4, 15, 100$ for an $O=3$ filter defined by Eq.~\eqref{eqn:butterworth}.}
\label{fig:filter_resp}
\end{center}
\end{figure}

Using smoothing process based on the DCT in Eq.~\eqref{eqn:smooth_DCT}, we can write
\begin{eqnarray}
\norm{\mathbf{D}_1\tilde{\mathbf{i}}}^2
&=&\mathbf{i}^T \mathbf{C}\mathbf{D}_{\rm filter}\mathbf{C}^T\mathbf{D}_1^T\mathbf{D}_1\mathbf{C}\mathbf{D}_{\rm filter}\mathbf{C}^T\mathbf{i} \nonumber \\
&=&\mathbf{i}^T\mathbf{C}\mathbf{D}_{\rm filter}\mathbf{\Lambda}_{D_1^TD_1}\mathbf{D}_{\rm filter}\mathbf{C}^T\mathbf{i} \nonumber\\
&=&\norm{\mathbf{\Lambda}_{D_1^TD_1}^{1/2}\mathbf{D}_{\rm filter}\mathbf{C}^T\mathbf{i}}^2
\label{eqn:sum_d1_smooth_DCT}
\end{eqnarray}
where we used the eigendecomposition $\mathbf{D}_1^T\mathbf{D}_1=\mathbf{C}\mathbf{\Lambda}_{D_1^TD_1}\mathbf{C}^T$ and $\mathbf{\Lambda}_{D_1^TD_1}$ is a diagonal matrix of non-negative eigenvalues, $\text{eig}_k$, shown in Eq.~\eqref{eqn:eval_DCT2}; $\mathbf{C}^T\mathbf{C}=\mathbf{I}$ since $\mathbf{C}$ is an orthogonal matrix. Eq.~\eqref{eqn:sum_d1_smooth_DCT} is a key result. It expresses the final product which is the sum of the square of the derivative of the smoothed data in terms of the end-to-end matrix operation. We note that $\mathbf{\Lambda}_{D_1^TD_1}^{1/2}$ is a diagonal matrix whose diagonal represents a high pass profile such that $\mathbf{\Lambda}_{D_1^TD_1}^{1/2}\mathbf{D}_{\rm filter}$ represents a band pass filter with the peak response at approximately $k_c$ as shown in Fig.~\ref{fig:filter_resp}. The norm (subscript $._2$ is not shown for brevity)  of the smoothed data is the structure parameter, $\sqrt{\Sigma(d/dt)^2}$,
\begin{eqnarray}
\norm{\mathbf{D}_1\tilde{\mathbf{i}}}=\norm{\mathbf{\Lambda}_{D_1^TD_1}^{1/2}\mathbf{D}_{\rm filter}\mathbf{C}^T\left(\tilde{\mathbf{i}}+\mathbf{\Delta_i}\right)}. 
\label{eqn:norm_res}
\end{eqnarray} 
Finally, the uncertainty is 
\begin{eqnarray}
\norm{\mathbf{D}_1\mathbf{S}\mathbf{\Delta}_\mathbf{i}}=\norm{\mathbf{\Lambda}_{D_1^TD_1}^{1/2}\mathbf{D}_{\rm filter}\mathbf{C}^T\mathbf{\Delta_i}}.
\label{eqn:norm_del}
\end{eqnarray}
Expressing the 2-norm in the spectral domain as we have done above has an advantage in visualization. For example, $\mathbf{C}^T\mathbf{i}$ is the spectrum of $I(t)$. Pre-multiplying that quantity by $\mathbf{\Lambda}_{D_1^TD_1}^{1/2}\mathbf{D}_{\rm filter}$ filters the spectrum. Finally, taking the norm measures the length of the resulting vector.

\section{Result: Numerical Example} \label{sec:result}

\subsection{Selection of Filter Parameters} \label{sec:numex_filt}
\begin{figure}[htb]
\begin{center}
\noindent
  \includegraphics[width=3.25in]{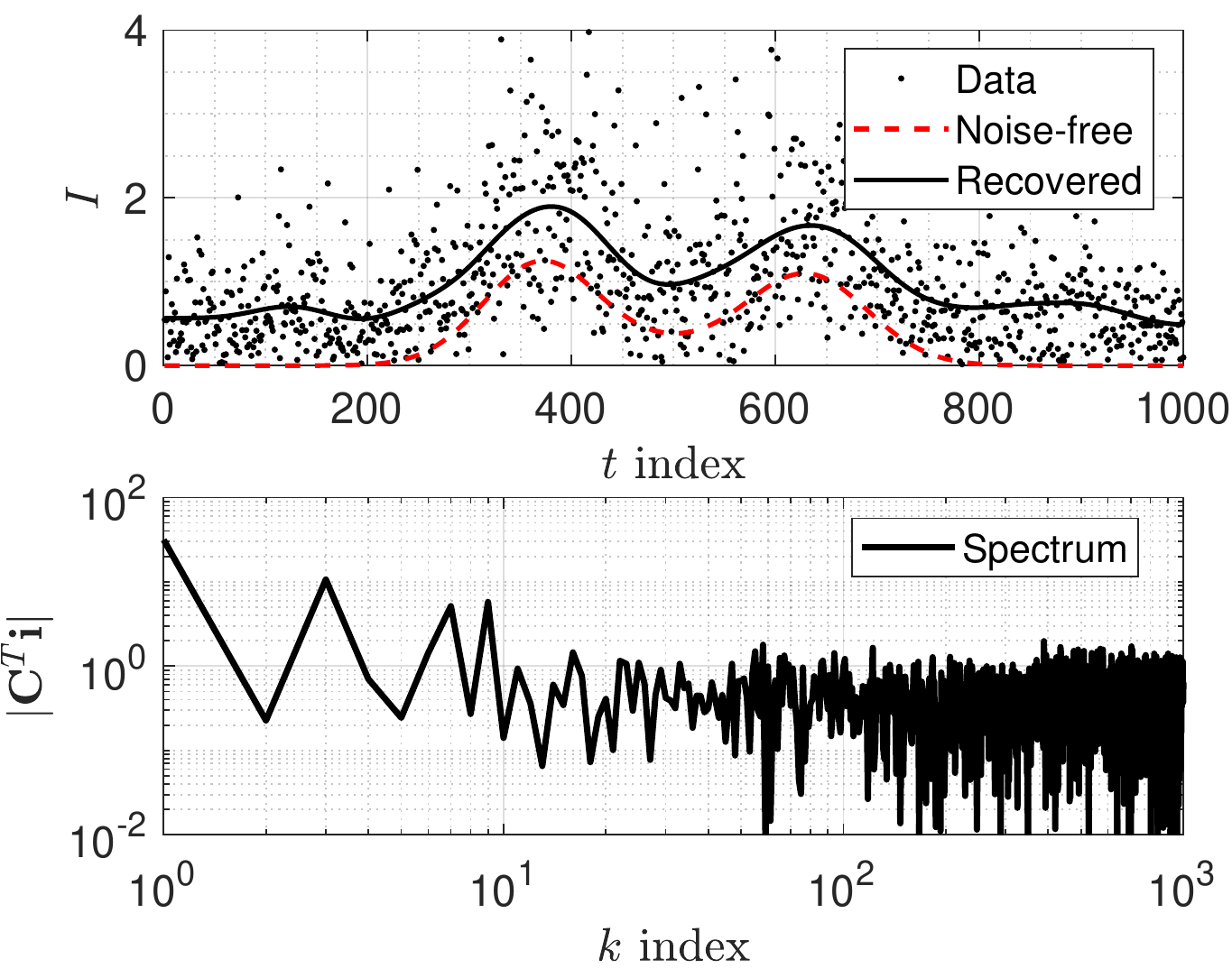}
\caption{Example noisy pulse (top) and the corresponding spectrum in discrete cosine (bottom). The underlying pulse consists of two unity amplitude Gaussians. The solid black curve in the top plot represents the recovered intensity with filter parameters $O=3,~k_c=15$.}
\label{fig:I_and_spec}
\end{center}
\end{figure}

We exemplify the method discussed in  \secref{sec:unc_def} by demonstrating the selection of filter parameters to correctly recover the underlying $I(t)$ structure in the presence of noise. Fig.~\ref{fig:I_and_spec} (top) shows 1000 noisy $I=|X|^2+|Y|^2$ data points with an underlying double pulse as shown, representing total power $I$ calculated from perpendicular $X$ and $Y$ linearly polarized voltage components. The noise-free component of $X$ consists of two Gaussian pulses delayed by 370 and 630 units with standard deviations of 80 and 85, respectively. The amplitude is normalized to the pulse peak. The noise-free amplitude of the $Y$ component is taken as $0.5X$. The structure parameter of the noise-free pulse is $\norm{\mathbf{D}_1\mathbf{i}_{\text{noiseless}}}=0.18$ as computed by the 2-norm of the first derivative. The noise in the complex voltages $X,~Y$ are independent and identically distributed Gaussian noise with zero mean and standard deviation $\sigma_{\Re(X)}=\sigma_{\Im(X)}=\sigma_{\Re(Y)}=\sigma_{\Im(Y)}=0.4$. The DCT spectrum of the signal is depicted in Fig.~\ref{fig:I_and_spec} (bottom). It shows that the information occupies up to a $k$ index of 10 to 20, beyond which the noise dominates.  Therefore, we expect to select a low-pass filter cutoff between $k_c$ of 10 to 20. 

\begin{figure}[htb]
\begin{center}
\noindent
  \includegraphics[width=3.25in]{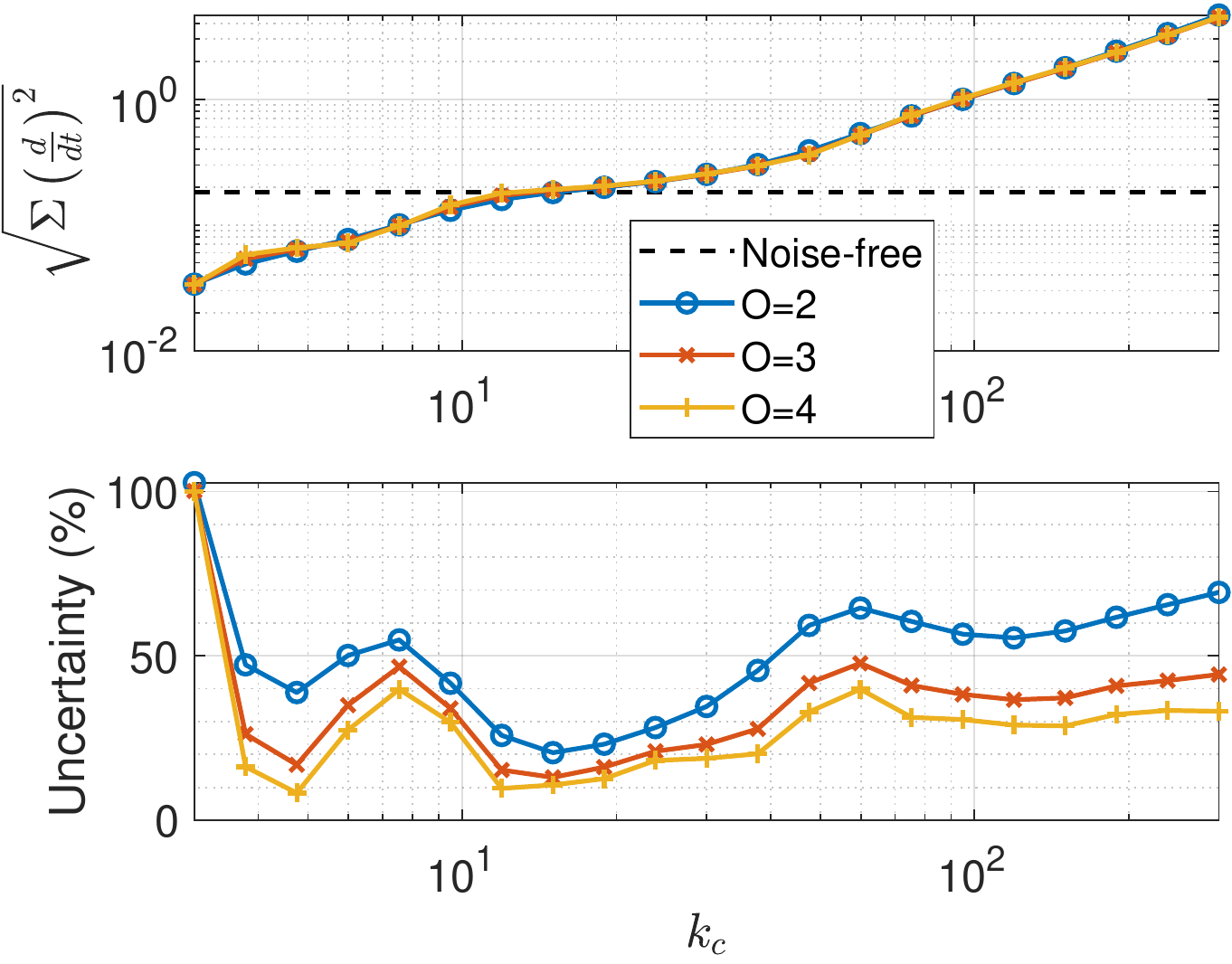}
\caption{Recovered structure parameter (top) and uncertainty (bottom) of the structure parameter for $k_c$ swept from 3 to 300 and filter order 2, 3, 4 in Eq.~\eqref{eqn:butterworth}.}
\label{fig:unc_filter}
\end{center}
\end{figure}

The $k_c$ estimate based on the DCT spectrum is confirmed by the calculated structure parameter and the uncertainty shown in Fig.~\ref{fig:unc_filter} where the low-pass filter cutoff $k_c$ was swept from 3 to 300 for filter order 2, 3, 4 in Eq.~\eqref{eqn:butterworth} to explore the effects of under- and over-filtering.  The structure parameter was computed using Eq.~\eqref{eqn:norm_res} and the associated uncertainty was calculated using Eq.~\eqref{eqn:rel_error}. The structure parameter obtained from the smoothed noisy data matches the known noise-free value of 0.18 for $k_c$ between 10 to 20, which agrees with that suggested by the DCT spectrum. This also coincides with the range of $k_c$ values for which the uncertainty is minimized. We also note that the structure parameter becomes rapidly underestimated and overestimated as the $k_c$ is underestimated (over-smoothing) and overestimated (under-smoothing), respectively. The uncertainty minima at $k_c$ of approximately 5 for filter orders 3 and 4 in Fig.~\ref{fig:unc_filter} correspond to over-smoothing which may be readily identified from the DCT spectrum in Fig.~\ref{fig:I_and_spec} and confirmed by the recovered structure parameter. Proper determination of $k_c$ is therefore critical, and we demonstrate that the DCT spectrum is a very effective tool for that purpose.

\subsection{Robustness with respect to $\Delta$\,DM} \label{sec:numex_DM}

The previous sub-section operated on one-dimensional time-series data, which assumes that de-dispersion has already been performed.  However, in reality, the dispersion measure is not known a priori (otherwise the problem would already be solved!) and hence it is necessary to consider whether de-dispersing across different trial DM values could affect the parameter choice and hence final results. In this sub-section, we simulate the effects of dispersion and de-dispersion on the double pulse already considered.

In the frequency domain, the net effect of dispersion and de-dispersion may be described as a multiplicative factor
\begin{eqnarray}
e^{-j(k(\nu)-\tilde{k}(\nu))L} = e^{-j(2\pi)K_{\rm DM}\frac{-\Delta\text{DM}\text[\pccc]}{\nu[\text{MHz}]}}. 
\label{eqn:disp-dedisp}
\end{eqnarray}
The units in Eq.~\eqref{eqn:disp-dedisp} follow the convention described in~\citet{Wilson2009}; $L$ is the distance to the source and $k(\nu)$ and $\tilde{k}(\nu)$ are the wavenumber and the wavenumber estimate, respectively; the difference between the actual DM and the DM estimate is $\Delta\text{DM}=\text{DM}-\tilde{\text{DM}}$.
Throughout, we define the delay constant $K_{\rm DM}$ to be $(1/0.241) \cdot 10^9 \approx 4.149 \cdot 10^9$\,MHz\,cm$^{3}$\,pc$^{-1}$.
Hence, the effects of dispersion and de-dispersion may be simulated by multiplying Eq.~\eqref{eqn:disp-dedisp} with the Fourier transform (FT) of the noisy complex voltages, followed by inverse FT of the product.

\begin{figure}[htb]
\begin{center}
  \includegraphics[width=3.25in]{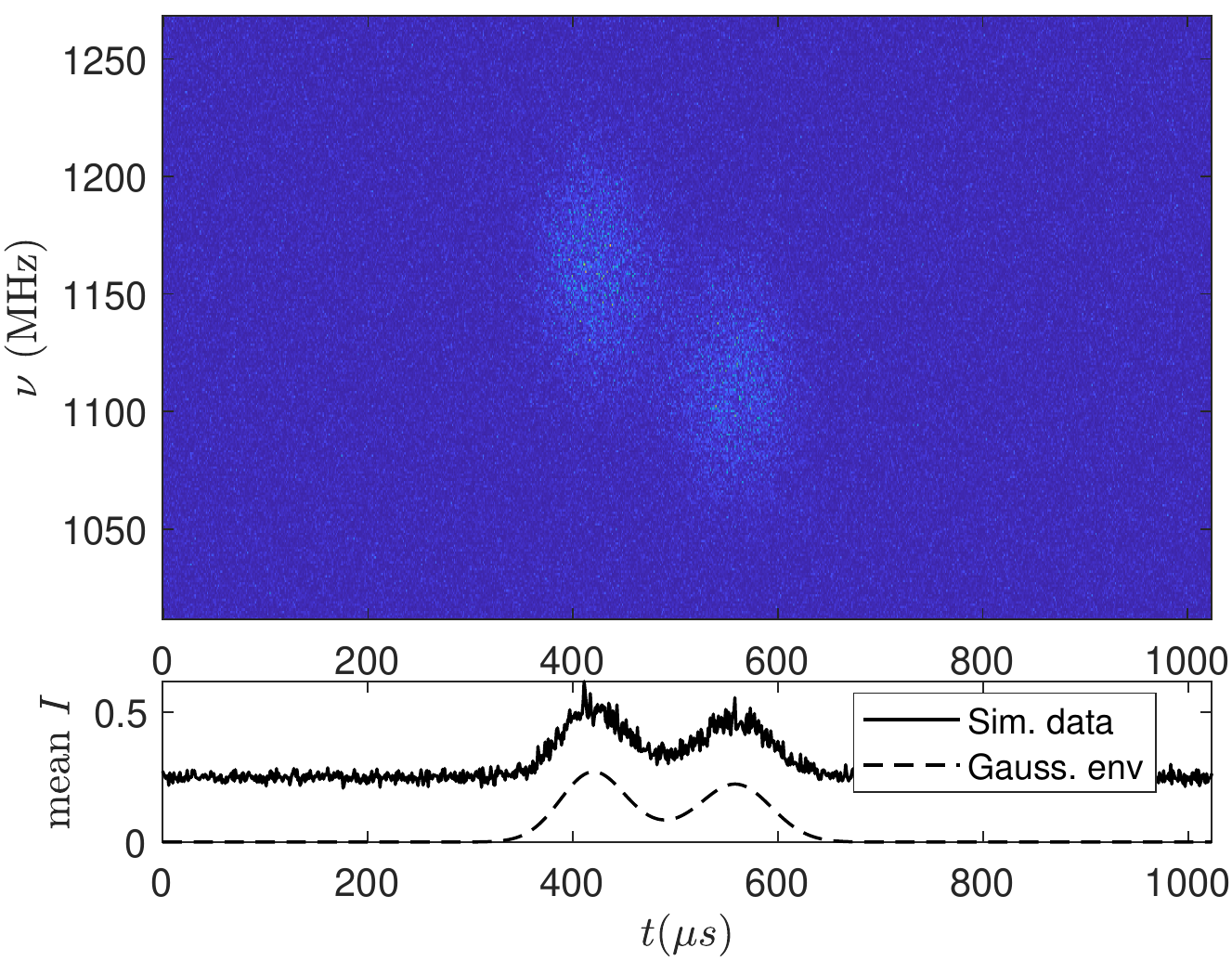}
\caption{The frequency-time simulated frequency down-drifting data at $\Delta\text{DM}=0$ (top). The bottom figure shows the mean intensity over frequency of the noisy data (solid line) and the Gaussian envelope (dashed line).}
\label{fig:FT_DM0}
\end{center}
\end{figure}

As an example, we took a $256\times1024$ matrix of complex white noise with $\sigma_{\rm re}=\sigma_{\rm im}=1/\sqrt{2}$. This noise matrix was multiplied by two 2-D Gaussian real envelopes, then complex Gaussian noise with $\sigma_{\rm re}=\sigma_{\rm im}=1/2\sqrt{2}$ was added. The result shown in Fig.~\ref{fig:FT_DM0} is the simulated data at $\Delta\text{DM}=0$. We chose the time delays between the components to mimic the `sad trombone' effect, i.e.\ the frequency down-drift of burst components with time, often seen in repeating FRBs \citep[e.g.][]{Hessels_2019}.

\begin{figure}[htb]
\begin{center}
  \includegraphics[width=3.25in]{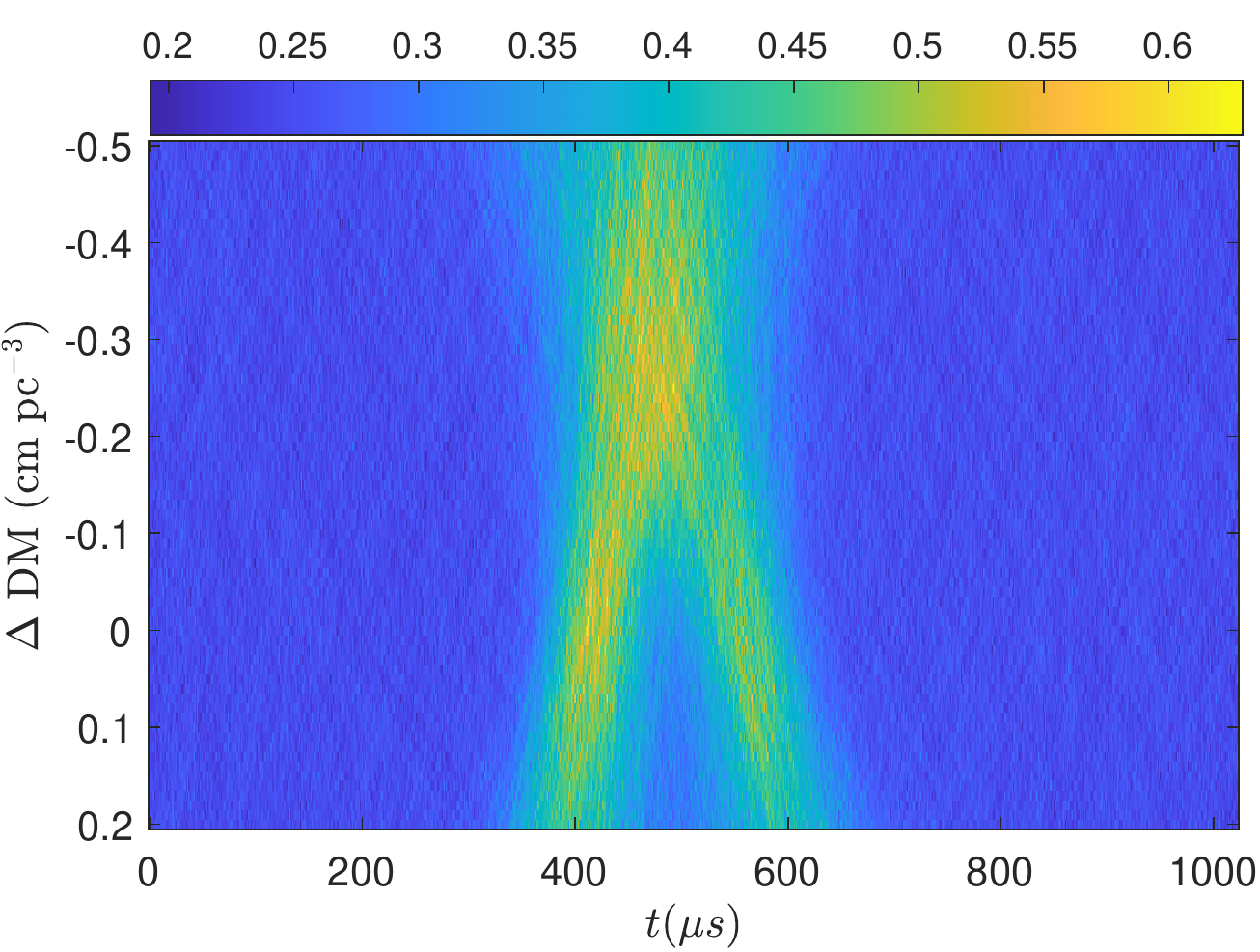}
\caption{Intensities vs.\ time and $\Delta \text{DM}$ of simulated frequency down-drifting data before filtering.}
\label{fig:I_DM_noF}
\end{center}
\end{figure}

We applied $\Delta\text{DM}$ from $-0.5$ to $0.2\, \text{pc} \,\text{cm}^{-3}$ in $0.01\,\text{pc}\, \text{cm}^{-3}$ steps. The resulting noisy mean intensities vs.\ time over $\Delta\text{DM}$ are shown in Fig.~\ref{fig:I_DM_noF}. The intensity of the noise-free data was simulated by applying the corresponding group delay $\tau_d=4.149\times10^3\Delta \text{DM} [\text{pc} \, \text{cm}^{-3}]/f_{\rm MHz}^2$ to the intensity envelope. To filter the noise in Fig.~\ref{fig:I_DM_noF}, we applied the same strategy discussed in \secref{sec:numex_filt}. We performed DCT of the frequency-time  $256\times1024$ data shown Fig.~\ref{fig:FT_DM0} and then visually inspected the resulting DCT spectrum (not shown). The transition between information and noise occurred at $k$ index of approximately 35. Therefore, we set $k_c=35$ and $O=3$ was kept the same as the previous section.

\begin{figure*}
\centering
  \includegraphics[width=6in]{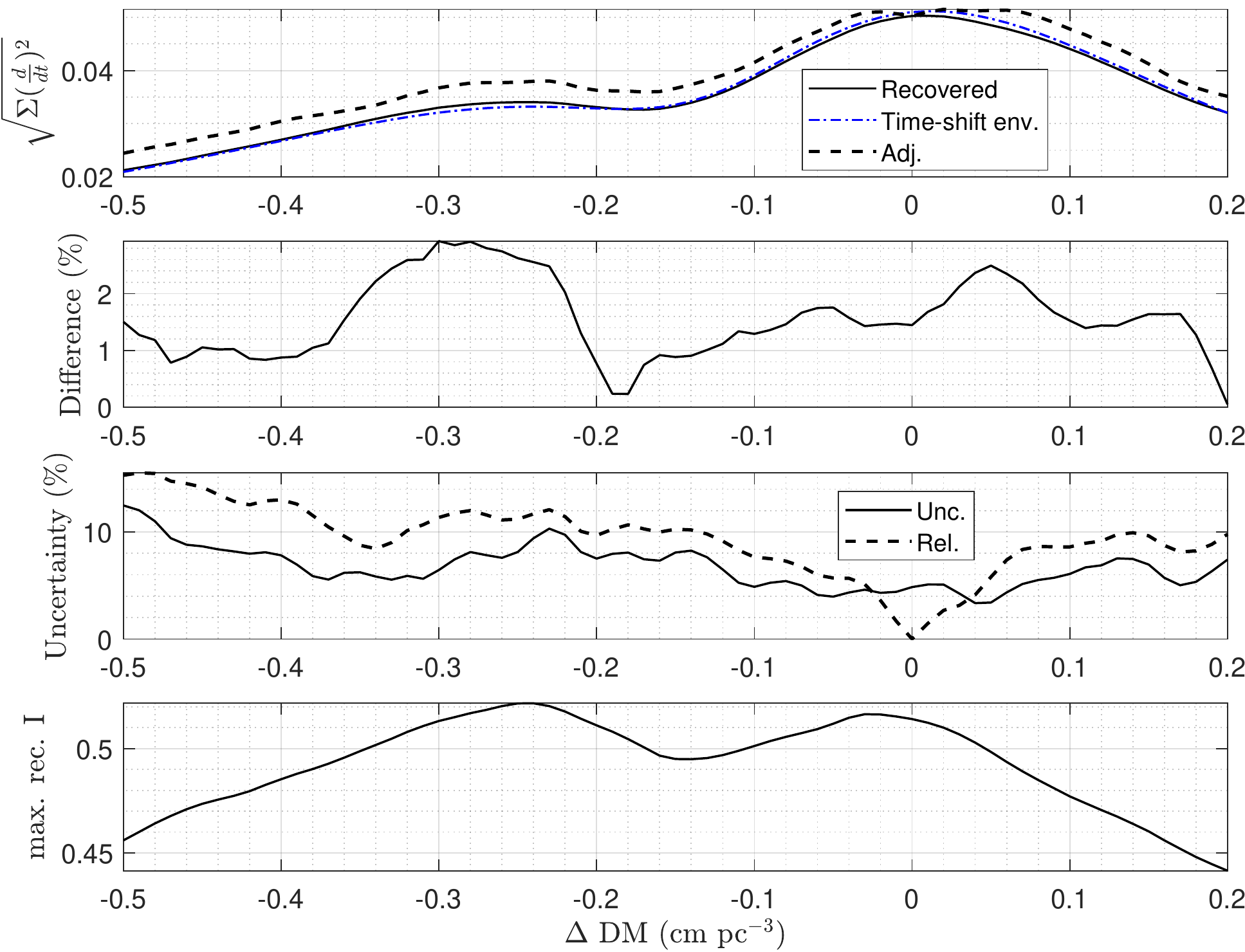}
\caption{Calculated structure parameter (top), difference between the structure parameter of the recovered signal and the original time-delayed envelope (second row), uncertainty of the recovered structure parameter (third row), maximum recovered intensity (bottom), of simulated frequency down-drift data.
}
\label{fig:Stuct_DM}
\end{figure*}

Fig.~\ref{fig:Stuct_DM}  summarizes the robustness of the method we propose in this paper with respect to $\Delta\text{DM}$. Fig.~\ref{fig:Stuct_DM} (top) shows the comparison between the recovered (from the noisy data) and the known structure parameters (from the noise-free intensity envelope that was time shifted as per the group delay at each frequency).
The difference between the two are plotted in the second row. The uncertainty computed using Eq.~\eqref{eqn:rel_error} is shown in the third row. Finally, the maximum recovered intensity is shown at the bottom of  Fig.~\ref{fig:Stuct_DM}. From this figure, we infer the following. The filter recovers the correct structure parameters to within a few percent for all trial $\Delta\text{DM}$s. The recovered highest structure parameter at $\Delta\text{DM}\approx 0$ and the maximum intensity at $\Delta\text{DM}\approx -0.26$\,\pccc\ are consistent with the noise-free intensity envelope. For the selected filter, the uncertainty of $\sim5$\% to 10\% in the measured structure is comparable across $\Delta\text{DM}$s. At $\Delta\text{DM}=0$, where the structure is maximised but the intensity is lower than the highest peak, the uncertainty appears slightly lower.

\begin{figure}
    \centering
    \includegraphics[width=3.25in]{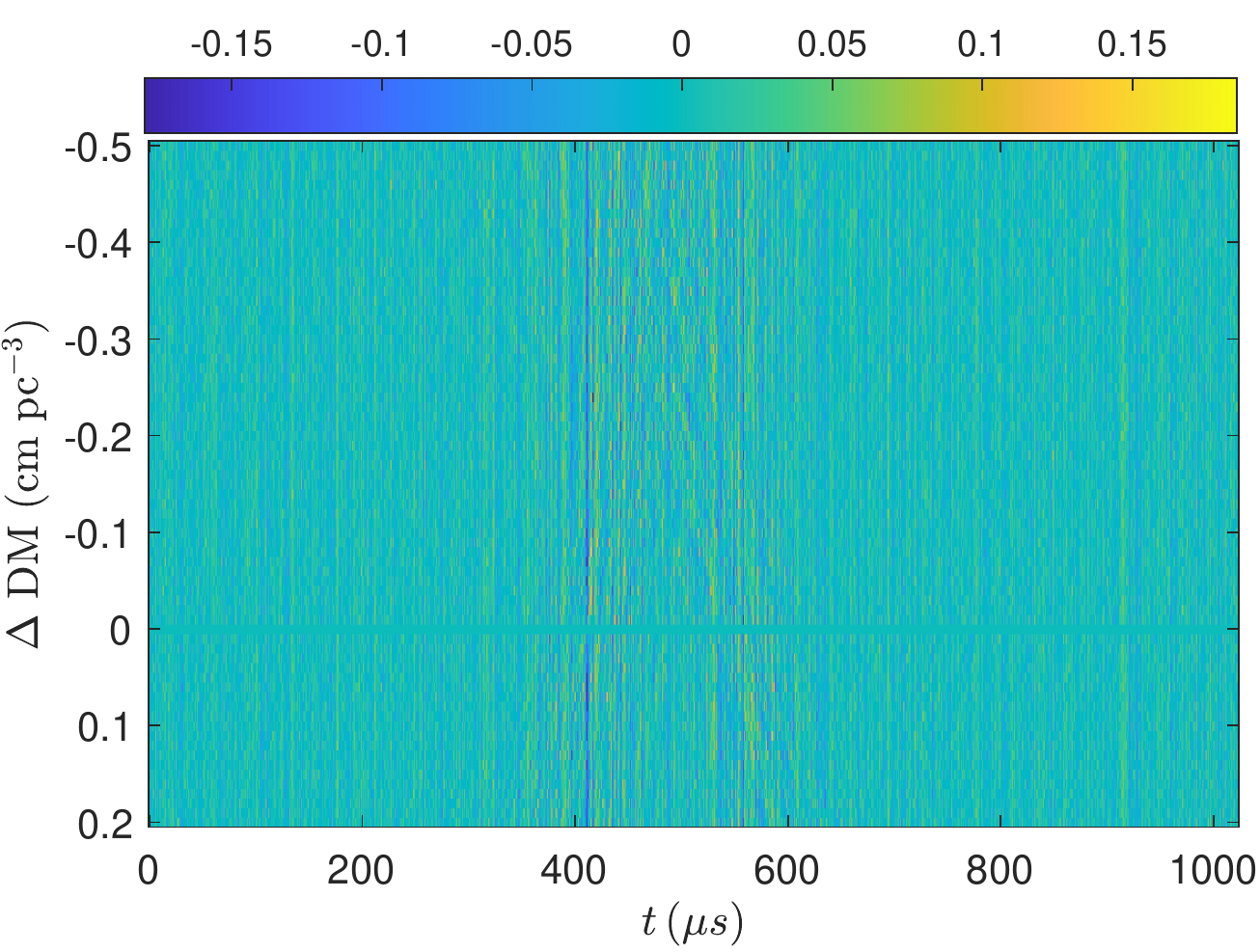}
    \caption{The vector $\mathbf{D}_1\mathbf{S}\left( \mathbf{\Delta}_{\Delta {\rm DM}=0} - \mathbf{\Delta}_{\Delta {\rm DM}} \right)$ from Eq.~\eqref{eqn:rel_dmerror}, calculated for simulated frequency down-drifting data. }
    \label{fig:relative_noise_example}
\end{figure}

To derive an uncertainty in $\Delta \text{DM}$, we take the noise estimates $\Delta_{\mathbf{i}}$ at the structure-maximised value $\Delta$DM=0, and subtract the noise estimates at other values of $\Delta$DM. This is shown in Figure~\ref{fig:relative_noise_example}. The magnitude of the relative errors as per Eq.~\eqref{eqn:rel_dmerror} are given in Fig.~\ref{fig:Stuct_DM}. For values of $\Delta$DM near $0$, the relative uncertainty (dashed line in third panel from top) is smaller than the absolute uncertainty (solid line) due to strong correlations in the filtered noise. At large values of $\Delta$DM however, the noise becomes uncorrelated, and so is approximately $\sqrt{2}$ times the absolute value.

To derive error bounds for $\Delta$DM, we add the error to the structure as per the right-hand-side of Eq.~\eqref{eqn:add_error}, also shown in Fig.~\ref{fig:Stuct_DM}. The uncertainty in $\Delta$DM corresponds to those regions where the relative noise plus the intrinsic structure exceeds that of the structure at $\Delta$DM$=0$, i.e.\ which can not be excluded by Eq.~(\eqref{eqn:add_error}). Thus we find $\Delta$DM$=0_{-0.04}^{+0.08}$\,\pccc.

\section{Result: Application to FRB data} \label{sec:apply}

The above methods can be applied to determine the correct DM of an FRB, and the uncertainty thereon. Dedispersing an FRB for many trial DMs produces voltage time series that can be converted into intensity time-series $\mathbf{i}$. The procedure of Sec.~\secref{sec:method} then produces a corresponding measure of FRB structure and its uncertainty for each and every trial DM. This then allows the trial DM that maximizes structure to be identified, and an error in DM assigned to include the range of trial DMs with structure consistent with the maximum value, given the calculated structure error.

\subsection{FRB181112} \label{sec:181112}
\begin{figure}[htb]
\begin{center}
\noindent
    \includegraphics[width=3.0in]{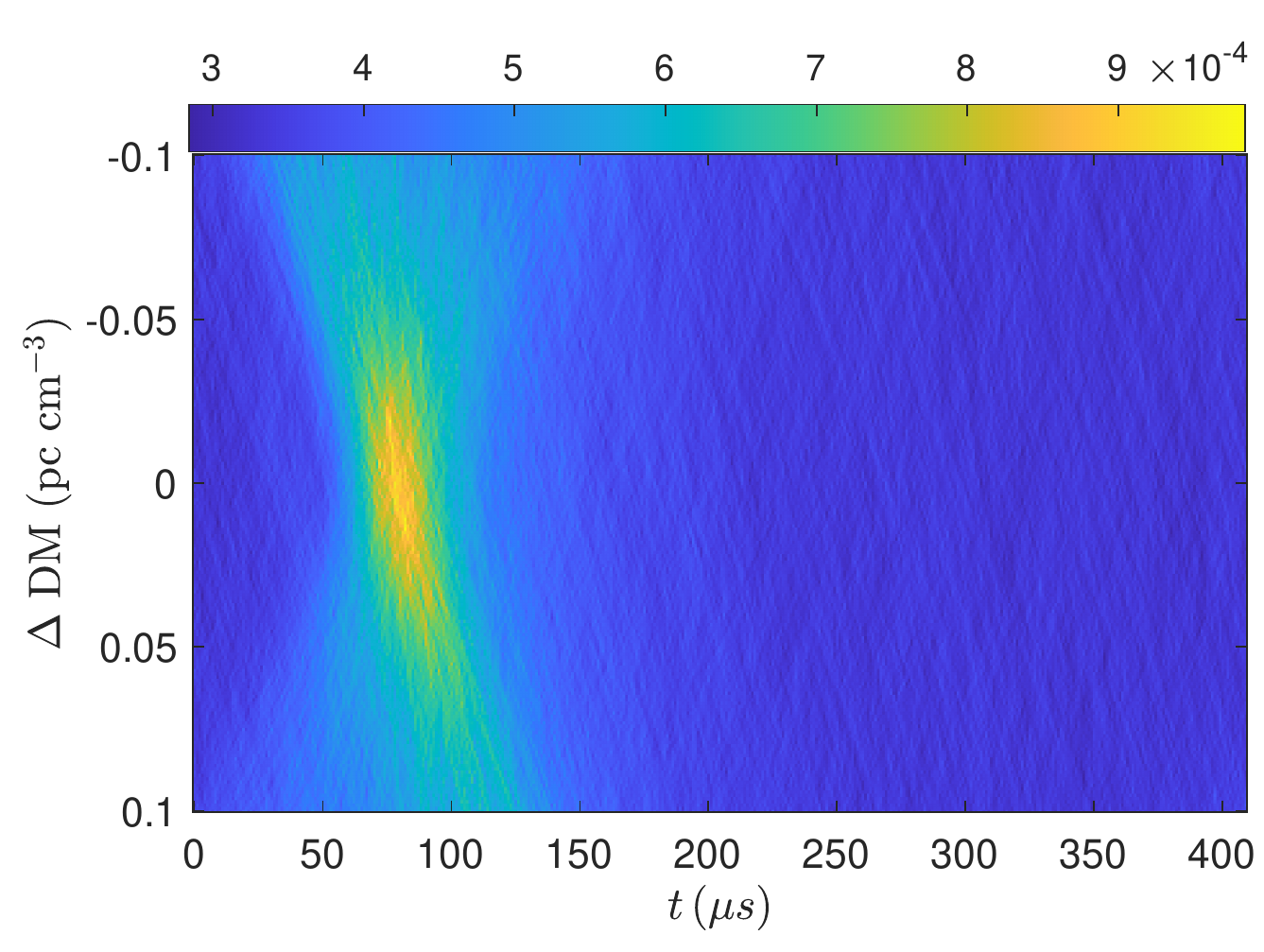}
\caption{Intensity of FRB181112 vs.\ time and $\Delta$ DM in pc\,cm$^{-3}$ from the DM that produces maximum S/N.}
\label{fig:FRB_DM_t}
\end{center}
\end{figure}

\begin{figure}[htb]
\begin{center}
\noindent
  \includegraphics[width=2.75in]{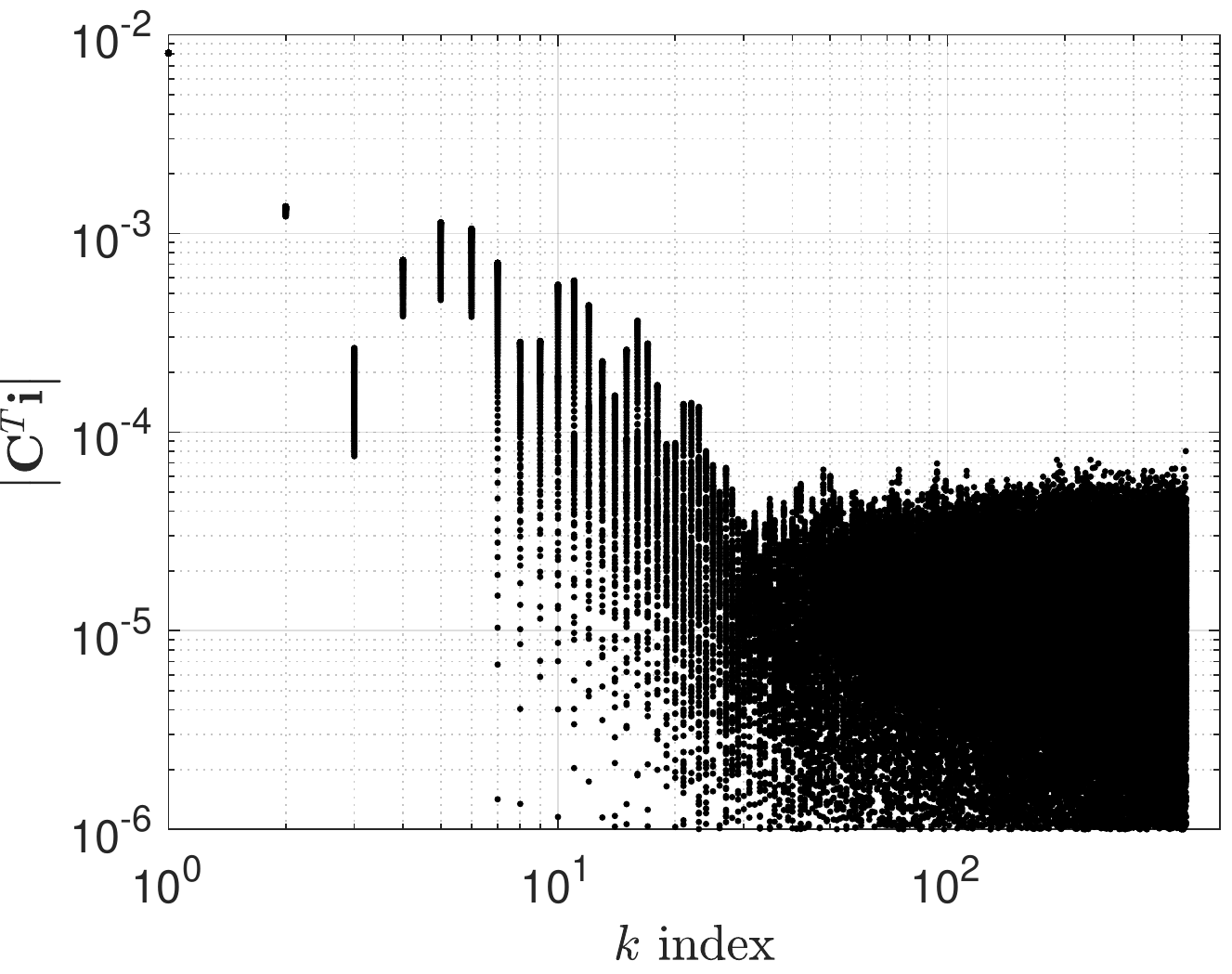}
\caption{Spectrum of FRB~181112 in discrete cosine domain shown in absolute values for all $\Delta$DM.}
\label{fig:FRB_spec}
\end{center}
\end{figure}

We apply our method to FRB181112~\citep{cho_2020ApJ...891L..38C} data averaged to 1\,$\mu$s. Fig.~\ref{fig:FRB_DM_t} shows the raw intensity data as a function of dispersion measure, away from the nominal value of 589.265\,pc\,cm$^{-3}$ that produces the maximum signal to noise (S/N) ratio. The spectrum of the intensity time series was obtained by the multiplication with the discrete cosine matrix discussed in Sec.~\ref{sec:smooth_DCT}. This is shown in Fig.~\ref{fig:FRB_spec}. We observe that the information occupies up to $k=30$ index, after which it is dominated by noise. Based on this information, we selected the filter cutoff of $k_c=30$ and chose a $O=3$ filter as per Eq.~\eqref{eqn:butterworth}.

\begin{figure}[htb]
\begin{center}
\noindent
  \includegraphics[width=3.0in]{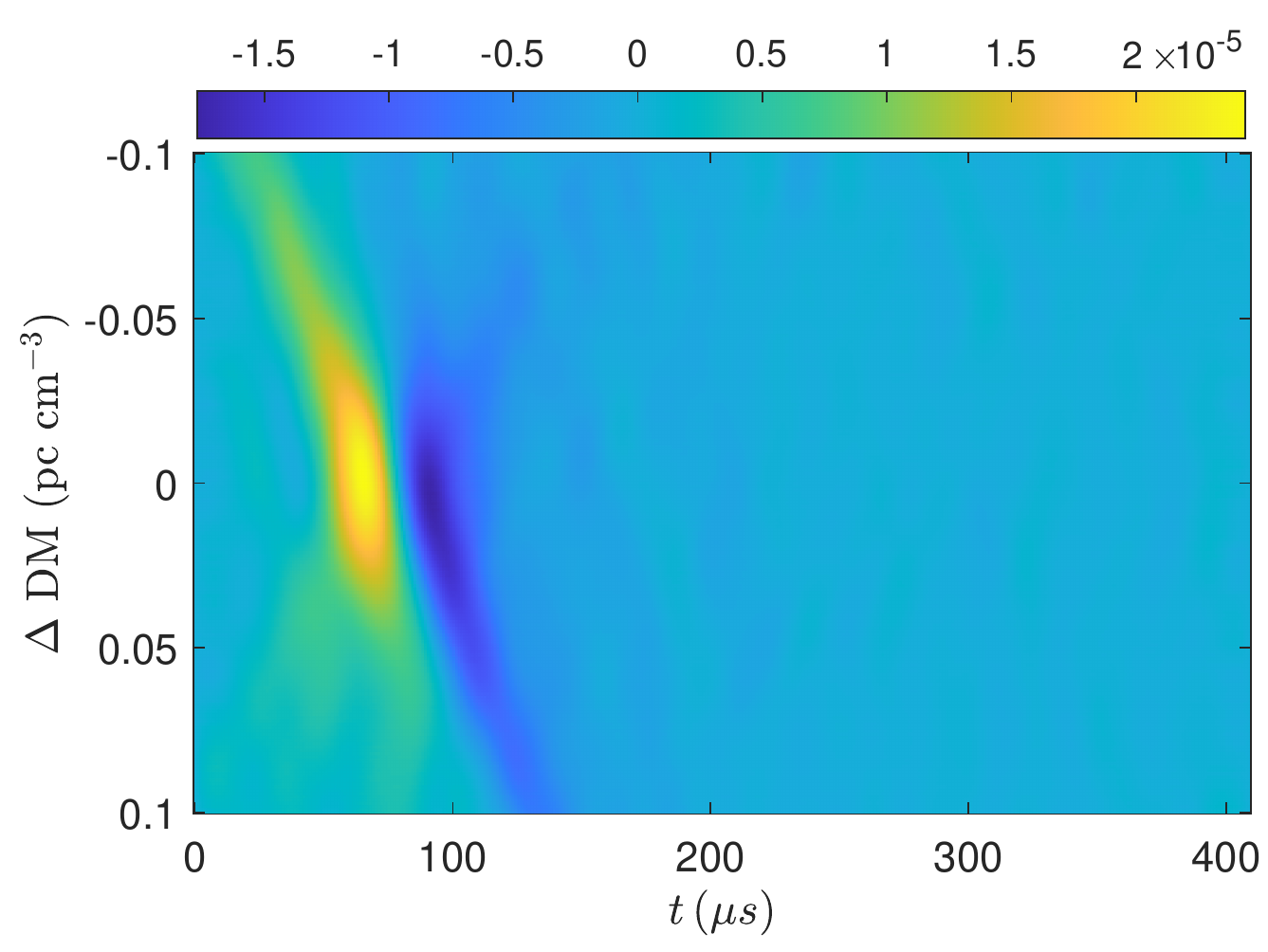}
\caption{First derivative of the intensity of FRB181112 vs.\ time and $\Delta$DM after smoothing using $k_c=30$ and $O=3$ filter.}
\label{fig:FRB_D1_smooth}
\end{center}
\end{figure}

\begin{figure}[htb]
\begin{center}
\noindent
  \includegraphics[width=3.25in]{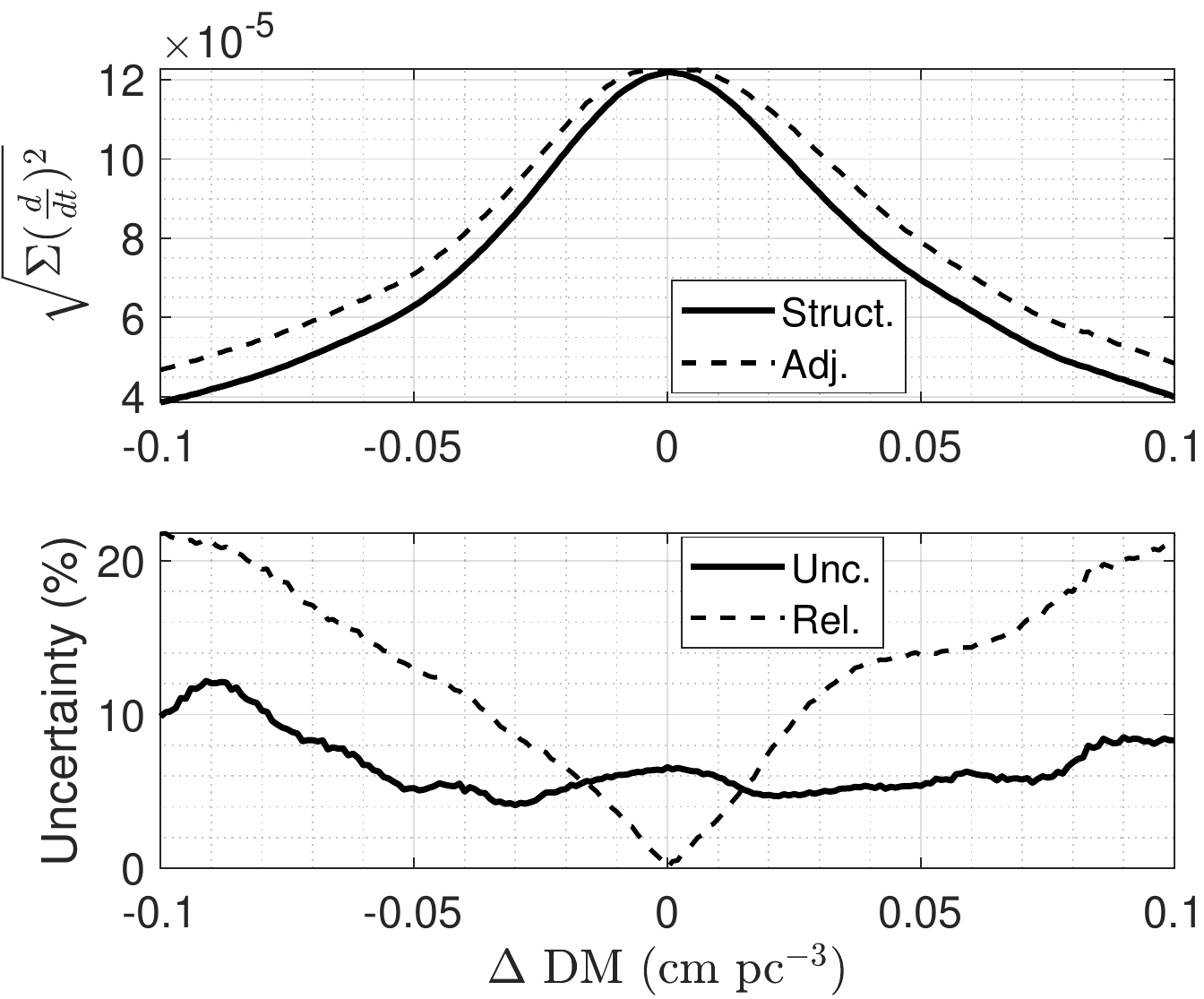}
\caption{Solid lines: Structure parameter of FRB181112 computed using $\sqrt{\Sigma(d/dt)^2}$ ($k_c=30, O=3$) (top), and the corresponding uncertainty (bottom). Dashed lines: 68\% upper limits on structure (top), and relative error (bottom), with respect to $\Delta$DM$=0$\,\pccc.}
\label{fig:FRB_struc}
\end{center}
\end{figure}

The first derivative of the smoothed data is reported in Fig.~\ref{fig:FRB_D1_smooth}. Finally, the structure parameter and the corresponding uncertainty are plotted in Fig.~\ref{fig:FRB_struc}. We see that the highest structure coincides with $\Delta$DM=0 which is the maximum S/N. The uncertainty has a trend that increases away from $\Delta {\rm DM}=0$, while the relative uncertainty compared to $\Delta{\rm DM}=0$ increases even more strongly. We find that allowing the structure to fluctuate upwards by one standard deviation of relative noise allows structure within $\Delta{\rm DM} = 0 \pm 0.011$\,\pccc\ to exceed the peak value of $12.2\times10^{-5}$ at $\Delta {\rm DM}=0$, i.e.\ our uncertainty in $\Delta {\rm DM}$ is $0.011$\,\pccc.

\subsection{FRB210117} \label{sec:210117}
\begin{figure}[tb]
\begin{center}
\noindent
  \includegraphics[width=3.0in]{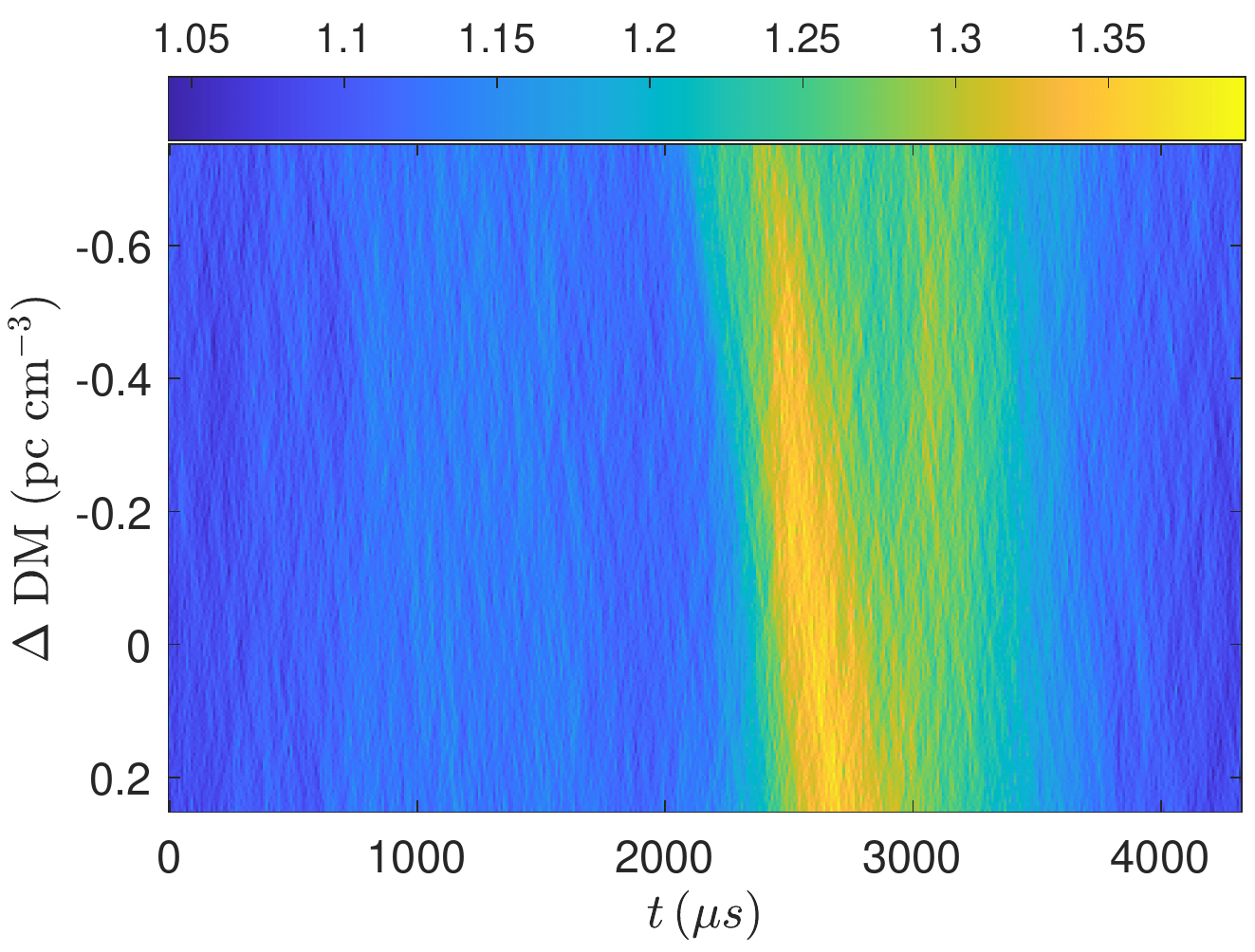}
\caption{Intensity before smoothing of FRB210117 vs.\ time and $\Delta$DM from the DM that produces maximum S/N.}
\label{fig:FRB210117_Raw}
\end{center}
\end{figure}

\begin{figure}[tb]
\begin{center}
\noindent
\includegraphics[width=3.0in]{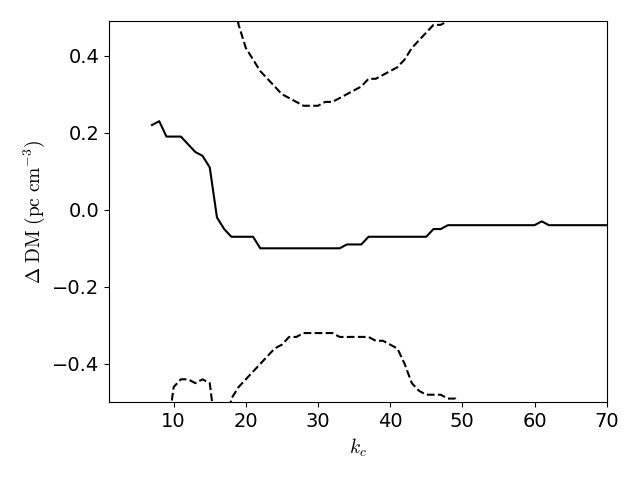}
\caption{Structure maximising $\Delta$DM (solid) and 68\% error bounds (dashed) for FRB210117 as a function of filter cutoff $k_c$ ($O=3$).}
\label{fig:FRB210117_kc_bounds}
\end{center}
\end{figure}

Next, we consider FRB210117 with time-domain data at 10\,$\mu$s resolution shown in Fig.~\ref{fig:FRB210117_Raw} for the indicated $\Delta$DM, relative to the nominal S/N maximizing value of 729.2\,pc\,cm$^{-3}$ \citep{Bhandari2022}. This FRB appears to have a secondary peak for $\Delta\text{DM}<0$. The spectrum of this FRB (not shown) suggests a spectral transition at a $k$ value between 25 and 35. This is shown in Fig.~\ref{fig:FRB210117_kc_bounds} where we plot the best-fitting $\Delta$DM and 68\% uncertainty bounds as per Eq.~\eqref{eqn:add_error} over filter cutoff $k_c$ (with $O=3$). It shows that $k_c\approx$25--35 offers consistent values of the structure-maximising $\Delta$DM with the narrowest error margin. Using lower values of $k_c$ results in over-smoothing, and thus a loss of signal structure; while using higher values of $k_c$ adds noise, increasing the random error in $\Delta$DM.

\begin{figure}[tb]
\begin{center}
\noindent
  \includegraphics[width=3.25in]{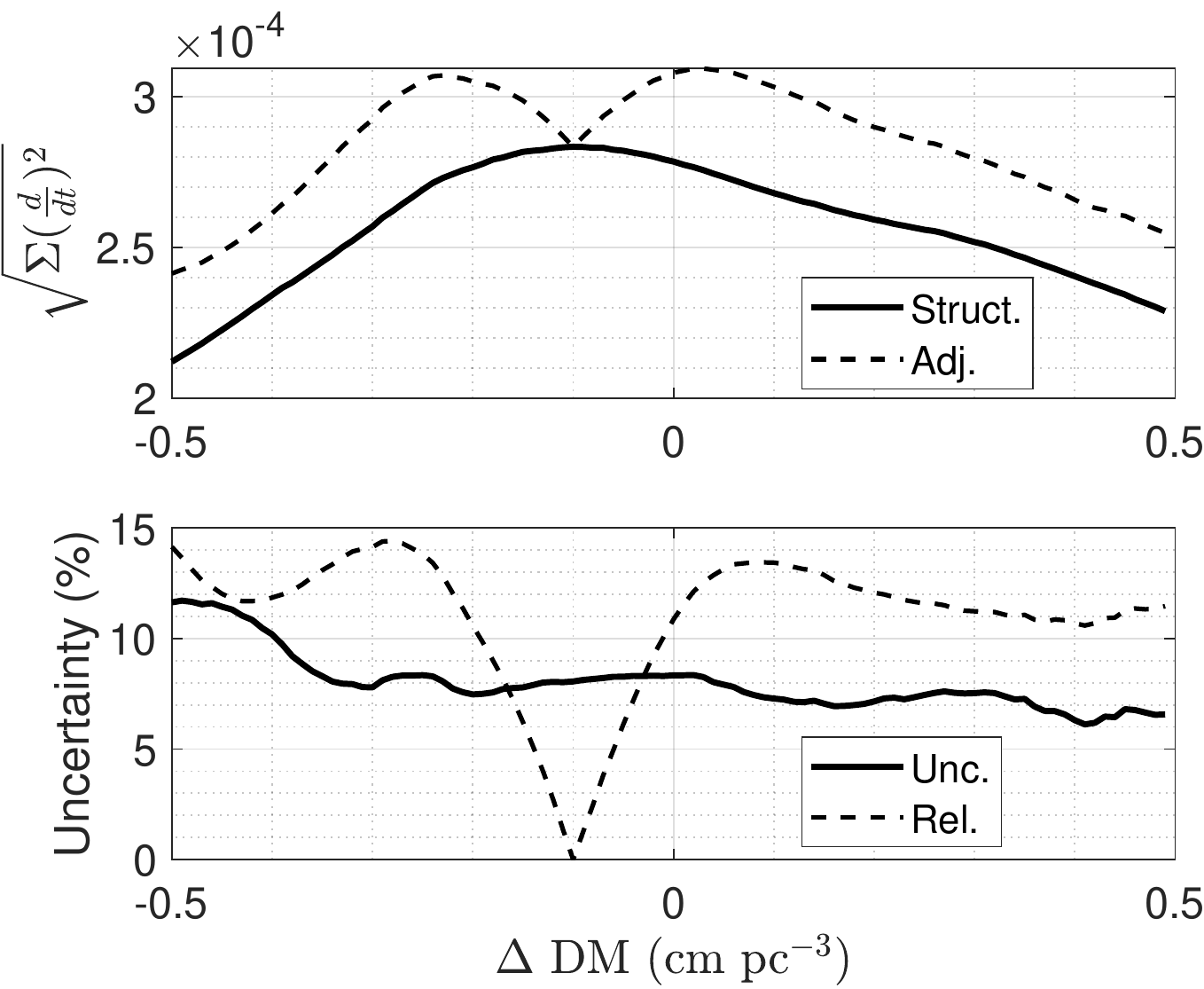}
\caption{Solid lines: Structure parameter of FRB210117 (top) and the corresponding uncertainty with $k_c=27$ and $O=3$ smoothing filter (bottom). Dashed lines: 68\% upper limit on structure (top), and relative uncertainty (bottom), with respect to $\Delta$DM$=-0.2$\,\pccc.}
\label{fig:FRB210117_struc_unc}
\end{center}
\end{figure}

The structure parameter and the associated uncertainty are plotted in Fig.~\ref{fig:FRB210117_struc_unc}. Using the same procedure as for FRB~20181112, we find a structure-maximising $\Delta\text{DM}=-0.1_{-0.23}^{+0.36}$\,\pccc. Thus, we cannot exclude that this FRB's apparent structure-maximising DM is significantly different from its signal-to-noise maximising DM.

\section{Conclusion} \label{sec:concl}
We demonstrated that computing the structure parameter using the 2-norm, $\sqrt{\Sigma (d/dt)^2}$, permits a direct calculation of the uncertainty of the resulting structure parameter. The uncertainty of the first derivative of the smoothed data is calculated by taking the difference between the structure parameter of the smoothed data and the structure parameter of the ``double-smoothed" data. The latter can be obtained easily by taking the smoothed data (before taking the derivative) and plugging it back as the input to the algorithm. We have also shown how to use this uncertainty for statistical tests for different intrinsic structure between two filtered signals.

This method was applied to FRB181112 and  FRB210117. In the case of FRB181112, the DM that produces the highest S/N is the same as that with the highest structure within an uncertainty in $\Delta$DM of $\pm0.011$\,\pccc. As for FRB210117, the highest structure occurred at a  $\text{DM}=729.1_{-0.23}^{+0.36}$\,\pccc, compared to the signal-to-noise maximising DM of $729.2$\,\pccc.

\begin{acknowledgments}
The Australian SKA Pathfinder is part of the Australia Telescope National Facility (https://ror.org/05qajvd42) which is managed by CSIRO. Operation of ASKAP is funded by the Australian Government with support from the National Collaborative Research Infrastructure Strategy. ASKAP uses the resources of the Pawsey Supercomputing Centre. Establishment of ASKAP, the Murchison Radio-astronomy Observatory and the Pawsey Supercomputing Centre are initiatives of the Australian Government, with support from the Government of Western Australia and the Science and Industry Endowment Fund. We acknowledge the Wajarri Yamatji people as the traditional owners of the Observatory site.
CWJ and MG acknowledge support from the Australian Government through the Australian Research Council's Discovery Projects funding scheme (project DP210102103). This work made use of \textsc{Scipy} \citep{SciPy2019}.
\end{acknowledgments}

\section*{Appendix} \label{sec:app}
We now briefly discuss other definitions of structure parameters: $\Sigma \left(d/dt\right)^2$~\citep{Hessels_2019}, $\Sigma \left|d/dt\right|$~\citep{Gajjar_2018}, and $\Sigma \left(d/dt\right)^4$~\citep{Josephy_2019}. We note that $\Sigma \left|d/dt\right|$ is the 1-norm of the vector derivative of the smoothed data~\citep[see Ch.~1]{Strang_LALD2019}. This means the triangle inequality is fulfilled, $\norm{\mathbf{x}+\mathbf{y}}_1\leq\norm{\mathbf{x}}_1+\norm{\mathbf{y}}_1$ and the upper bound is defined. However, the inner product is not defined in the 1-norm~\citep[see. Ch.~1]{Strang_LALD2019}. As a result, the corresponding lower bound cannot be shown to be that in Eq.~\eqref{eqn:D1_norm} for the 2-norm case\footnote{One can use the reverse triangle inequality to show $\norm{\mathbf{x}-\mathbf{y}}_1\geq\left|\norm{\mathbf{x}}_1-\norm{\mathbf{y}}_1\right|$, but this not the same as establishing the lower bound of $\norm{\mathbf{x}+\mathbf{y}}_1$.}. Furthermore, we are not able to relate the standard deviation estimate, such as Eq.~\eqref{eqn:norm_D1Si} in the 2-norm, to a corresponding statistical quantity in the 1-norm.

$\Sigma \left(d/dt\right)^2$ and $\Sigma \left(d/dt\right)^4$ do not conform to the norm definition because they are missing $\sqrt{}$ and $\sqrt[4]{}$ operators which would have made them 2-norm and 4-norm, respectively~\citep[see Ch.~7]{Laub_LA}. Therefore for these definitions, the triangle inequality does not apply and we are not able to establish the bounds of the summation of two vectors.

In \citet{Gajjar_2018}, the uncertainty of the structure parameter was inferred from the off-pulse system noise. This assumes that the noise during the FRB pulse is the same that of the system noise. However, it is entirely possible that the FRB pulse is noise-like and hence the on-pulse noise is different from the off-pulse noise. In fact, the noise statistics of the FRB is in itself an interesting subject for study, and therefore, it is desirable that the uncertainty be inferred from that same on-pulse data. 


\begin{thebibliography}{}
\expandafter\ifx\csname natexlab\endcsname\relax\def\natexlab#1{#1}\fi
\providecommand{\url}[1]{\href{#1}{#1}}
\providecommand{\dodoi}[1]{doi:~\href{http://doi.org/#1}{\nolinkurl{#1}}}
\providecommand{\doeprint}[1]{\href{http://ascl.net/#1}{\nolinkurl{http://ascl.net/#1}}}
\providecommand{\doarXiv}[1]{\href{https://arxiv.org/abs/#1}{\nolinkurl{https://arxiv.org/abs/#1}}}

\bibitem[{and B.~C.~Andersen {et~al.}(2019)and B.~C.~Andersen, Bandura,
  Bhardwaj, Boubel, Boyce, Boyle, Brar, Cassanelli, Chawla, Cubranic, Deng,
  Dobbs, Fandino, Fonseca, Gaensler, Gilbert, Giri, Good, Halpern, Hill,
  Hinshaw, Höfer, Josephy, Kaspi, Kothes, Landecker, Lang, Li, Lin, Masui,
  Mena-Parra, Merryfield, Mckinven, Michilli, Milutinovic, Naidu, Newburgh, Ng,
  Patel, Pen, Pinsonneault-Marotte, Pleunis, Rafiei-Ravandi, Rahman, Ransom,
  Renard, Scholz, Siegel, Singh, Smith, Stairs, Tendulkar, Tretyakov,
  Vanderlinde, Yadav, \& Zwaniga}]{Andersen2019}
and B.~C.~Andersen, Bandura, K., Bhardwaj, M., {et~al.} 2019, The Astrophysical
  Journal, 885, L24, \dodoi{10.3847/2041-8213/ab4a80}

\bibitem[{{Bhandari} {et~al.}(2022){Bhandari}, {Gordon}, {Scott}, {Marnoch},
  {Sridhar}, {Kumar}, {James}, {Qiu}, {Bannister}, {Deller}, {Eftekhari},
  {Fong}, {Glowacki}, {Prochaska}, {Ryder}, {Shannon}, \&
  {Simha}}]{Bhandari2022}
{Bhandari}, S., {Gordon}, A.~C., {Scott}, D.~R., {et~al.} 2022, arXiv e-prints,
  arXiv:2211.16790.
\newblock \doarXiv{2211.16790}

\bibitem[{Caleb {et~al.}(2020)Caleb, Stappers, Abbott, Barr, Bezuidenhout,
  Buchner, Burgay, Chen, Cognard, Driessen, Fender, Hilmarsson, Hoang, Horn,
  Jankowski, Kramer, Lorimer, Malenta, Morello, Pilia, Platts, Possenti,
  Rajwade, Ridolfi, Rhodes, Sanidas, Serylak, Spitler, Townsend, Weltman,
  Woudt, \& Wu}]{Caleb_10.1093/mnras/staa1791}
Caleb, M., Stappers, B.~W., Abbott, T.~D., {et~al.} 2020, Monthly Notices of
  the Royal Astronomical Society, 496, 4565, \dodoi{10.1093/mnras/staa1791}

\bibitem[{{Cho} {et~al.}(2020){Cho}, {Macquart}, {Shannon}, {Deller},
  {Morrison}, {Ekers}, {Bannister}, {Farah}, {Qiu}, {Sammons}, {Bailes},
  {Bhandari}, {Day}, {James}, {Phillips}, {Prochaska}, \&
  {Tuthill}}]{cho_2020ApJ...891L..38C}
{Cho}, H., {Macquart}, J.-P., {Shannon}, R.~M., {et~al.} 2020, \apjl, 891, L38,
  \dodoi{10.3847/2041-8213/ab7824}

\bibitem[{Eilers(2003)}]{Eilers_doi:10.1021/ac034173t}
Eilers, P. H.~C. 2003, Analytical Chemistry, 75, 3631,
  \dodoi{10.1021/ac034173t}

\bibitem[{Gajjar {et~al.}(2018)Gajjar, Siemion, Price, Law, Michilli, Hessels,
  Chatterjee, Archibald, Bower, Brinkman, Burke-Spolaor, Cordes, Croft,
  Enriquez, Foster, Gizani, Hellbourg, Isaacson, Kaspi, Lazio, Lebofsky, Lynch,
  MacMahon, McLaughlin, Ransom, Scholz, Seymour, Spitler, Tendulkar, Werthimer,
  \& Zhang}]{Gajjar_2018}
Gajjar, V., Siemion, A. P.~V., Price, D.~C., {et~al.} 2018, The Astrophysical
  Journal, 863, 2, \dodoi{10.3847/1538-4357/aad005}

\bibitem[{Hessels {et~al.}(2019)Hessels, Spitler, Seymour, Cordes, Michilli,
  Lynch, Gourdji, Archibald, Bassa, Bower, Chatterjee, Connor, Crawford,
  Deneva, Gajjar, Kaspi, Keimpema, Law, Marcote, McLaughlin, Paragi, Petroff,
  Ransom, Scholz, Stappers, \& Tendulkar}]{Hessels_2019}
Hessels, J. W.~T., Spitler, L.~G., Seymour, A.~D., {et~al.} 2019, The
  Astrophysical Journal, 876, L23, \dodoi{10.3847/2041-8213/ab13ae}

\bibitem[{Hilmarsson {et~al.}(2021)Hilmarsson, Spitler, Main, \&
  Li}]{Hilmarsson_10.1093/mnras/stab2936}
Hilmarsson, G.~H., Spitler, L.~G., Main, R.~A., \& Li, D.~Z. 2021, Monthly
  Notices of the Royal Astronomical Society, 508, 5354,
  \dodoi{10.1093/mnras/stab2936}

\bibitem[{Josephy {et~al.}(2019)Josephy, Chawla, Fonseca, Ng, Patel, Pleunis,
  Scholz, Andersen, Bandura, Bhardwaj, Boyce, Boyle, Brar, Cubranic, Dobbs,
  Gaensler, Gill, Giri, Good, Halpern, Hinshaw, Kaspi, Landecker, Lang, Lin,
  Masui, Mckinven, Mena-Parra, Merryfield, Michilli, Milutinovic, Naidu, Pen,
  Rafiei-Ravandi, Rahman, Ransom, Renard, Siegel, Smith, Stairs, Tendulkar,
  Vanderlinde, Yadav, \& Zwaniga}]{Josephy_2019}
Josephy, A., Chawla, P., Fonseca, E., {et~al.} 2019, The Astrophysical Journal,
  882, L18, \dodoi{10.3847/2041-8213/ab2c00}

\bibitem[{Laub(2005)}]{Laub_LA}
Laub, A.~J. 2005, Matrix analysis for scientists and engineers (Philadelphia,
  PA, USA: SIAM)

\bibitem[{{Lorimer} {et~al.}(2007){Lorimer}, {Bailes}, {McLaughlin},
  {Narkevic}, \& {Crawford}}]{Lorimer2007}
{Lorimer}, D.~R., {Bailes}, M., {McLaughlin}, M.~A., {Narkevic}, D.~J., \&
  {Crawford}, F. 2007, Science, 318, 777, \dodoi{10.1126/science.1147532}

\bibitem[{{Macquart} {et~al.}(2020){Macquart}, {Prochaska}, {McQuinn},
  {Bannister}, {Bhandari}, {Day}, {Deller}, {Ekers}, {James}, {Marnoch},
  {Os{\l}owski}, {Phillips}, {Ryder}, {Scott}, {Shannon}, \&
  {Tejos}}]{Macquart2020}
{Macquart}, J.~P., {Prochaska}, J.~X., {McQuinn}, M., {et~al.} 2020, \nat, 581,
  391, \dodoi{10.1038/s41586-020-2300-2}

\bibitem[{{Michilli} {et~al.}(2018){Michilli}, {Seymour}, {Hessels}, {Spitler},
  {Gajjar}, {Archibald}, {Bower}, {Chatterjee}, {Cordes}, {Gourdji}, {Heald},
  {Kaspi}, {Law}, {Sobey}, {Adams}, {Bassa}, {Bogdanov}, {Brinkman},
  {Demorest}, {Fernand ez}, {Hellbourg}, {Lazio}, {Lynch}, {Maddox}, {Marcote},
  {McLaughlin}, {Paragi}, {Ransom}, {Scholz}, {Siemion}, {Tendulkar}, {van
  Rooy}, {Wharton}, \& {Whitlow}}]{Michilli2018_121102}
{Michilli}, D., {Seymour}, A., {Hessels}, J.~W.~T., {et~al.} 2018, \nat, 553,
  182, \dodoi{10.1038/nature25149}

\bibitem[{Pilia {et~al.}(2020)Pilia, Burgay, Possenti, Ridolfi, Gajjar,
  Corongiu, Perrodin, Bernardi, Naldi, Pupillo, Ambrosino, Bianchi, Burtovoi,
  Casella, Casentini, Cecconi, Ferrigno, Fiori, Gendreau, Ghedina, Naletto,
  Nicastro, Ochner, Palazzi, Panessa, Papitto, Pittori, Rea, Castillo,
  Savchenko, Setti, Tavani, Trois, Trudu, Turatto, Ursi, Verrecchia, \&
  Zampieri}]{Pilia_2020}
Pilia, M., Burgay, M., Possenti, A., {et~al.} 2020, The Astrophysical Journal,
  896, L40, \dodoi{10.3847/2041-8213/ab96c0}

\bibitem[{{Platts} {et~al.}(2021){Platts}, {Caleb}, {Stappers}, {Main},
  {Weltman}, {Shock}, {Kramer}, {Bezuidenhout}, {Jankowski}, {Morello},
  {Possenti}, {Rajwade}, {Rhodes}, \& {Wu}}]{2021MNRAS.505.3041P}
{Platts}, E., {Caleb}, M., {Stappers}, B.~W., {et~al.} 2021, \mnras, 505, 3041,
  \dodoi{10.1093/mnras/stab1544}

\bibitem[{Stickel(2010)}]{STICKEL2010467}
Stickel, J.~J. 2010, Computers \& Chemical Engineering, 34, 467,
  \dodoi{https://doi.org/10.1016/j.compchemeng.2009.10.007}

\bibitem[{Strang(1999)}]{Strang_DCT}
Strang, G. 1999, SIAM Review, 41, 135, \dodoi{10.1137/S0036144598336745}

\bibitem[{Strang(2019)}]{Strang_LALD2019}
---. 2019, Linear Algebra and Learning from Data (Wellesley, MA, USA:
  Wellesley-Cambridge Press)

\bibitem[{Strang \& Nguyen(1997)}]{Strang_wavelet}
Strang, G., \& Nguyen, T.~Q. 1997, Wavelets and filter banks (Wellesley, MA,
  USA: Wellesley-Cambridge Press)

\bibitem[{{Virtanen} {et~al.}(2020){Virtanen}, {Gommers}, {Oliphant},
  {Haberland}, {Reddy}, {Cournapeau}, {Burovski}, {Peterson}, {Weckesser},
  {Bright}, {van der Walt}, {Brett}, {Wilson}, {Millman}, {Mayorov}, {Nelson},
  {Jones}, {Kern}, {Larson}, {Carey}, {Polat}, {Feng}, {Moore}, {VanderPlas},
  {Laxalde}, {Perktold}, {Cimrman}, {Henriksen}, {Quintero}, {Harris},
  {Archibald}, {Ribeiro}, {Pedregosa}, {van Mulbregt}, \& {SciPy 1. 0
  Contributors}}]{SciPy2019}
{Virtanen}, P., {Gommers}, R., {Oliphant}, T.~E., {et~al.} 2020, Nature
  Methods, 17, 261, \dodoi{10.1038/s41592-019-0686-2}

\bibitem[{Wilson {et~al.}(2009)Wilson, Rohlfs, \&
  H{\"u}ttemeister}]{Wilson2009}
Wilson, T.~L., Rohlfs, K., \& H{\"u}ttemeister, S. 2009, Electromagnetic Wave
  Propagation Fundamentals (Berlin, Heidelberg: Springer Berlin Heidelberg),
  19--37, \dodoi{10.1007/978-3-540-85122-6_2}

\bibitem[{{Zhao} {et~al.}(2021){Zhao}, {Zhang}, {Wang}, {Tu}, \&
  {Wang}}]{Zhao2021_DM_evolution}
{Zhao}, Z.~Y., {Zhang}, G.~Q., {Wang}, Y.~Y., {Tu}, Z.-L., \& {Wang}, F.~Y.
  2021, \apj, 907, 111, \dodoi{10.3847/1538-4357/abd321}

\end{thebibliography}

\end{document}